\documentclass[preprint,endfloats*,epsf,showpacs,amsmath,amssymb,12pt]{revtex4}

\usepackage{graphicx}
\usepackage{dcolumn}
\usepackage{bm}

\input epsf

\begin{document}
\newcommand{\ltwid}{\mathrel{\raise.3ex\hbox{$<$\kern-.75em\lower1ex\hbox{$\sim$}}}}
\newcommand{\gtwid}{\mathrel{\raise.3ex\hbox{$>$\kern-.75em\lower1ex\hbox{$\sim$}}}}

\title{The wave-vector power spectrum of the local \\
tunneling density of states: ripples in a d-wave sea}

\author{L.~Capriotti\email{caprio@kitp.ucsb.edu}}
\affiliation{Kavli Institute of Theoretical Physics\\
University of California, Santa Barbara, CA 93106-4030}

\author{D.J.~Scalapino\email{djs@vulcan2.physics.ucsb.edu} and 
R.D.~Sedgewick\email{rds@physics.ucsb.edu}}
\affiliation{Department of Physics, University of California\\ 
Santa Barbara, California 93106-9530}

\date{\today}

\begin{abstract}

A weak scattering potential imposed on a $CuO_2$ layer of a
cuprate superconductor modulates the local density of
states $N(x,\omega)$.  In recently reported experimental studies 
\cite{Hof02,How02,McE02},
scanning-tunneling maps of $N(x,\omega)$ have been Fourier transformed
to obtain a wave-vector power spectrum.  Here, for the case of a weak
scattering potential, we discuss the structure of this
power spectrum and its relationship to the quasi-particle
spectrum and the structure factor of the scattering
potential. Examples of quasi-particle interferences in normal
metals and $s$- and $d$-wave superconductors are discussed.
\end{abstract}

\maketitle



\section{Introduction}

A weak scattering potential imposed on the $CuO_2$ layer of a cuprate superconductor
creates ripples in the local tunneling 
density of states $N(x, \omega)$ due to quasi-particle
interference scattering.  It was suggested that scanning tunneling measurements of
the spatial and frequency structure of $N(x, \omega)$ could provide information on
the $k$ and $\omega$ dependence of the gap \cite{BFS93}. Recently, the introduction
of high-resolution Fourier-transform scanning tunneling microscopy
\cite{Hof02,How02,McE02} (FT-STM) has provided a powerful new technique for studying
this. In this approach, a BISCO crystal is cleaved exposing a $BiO_2$ layer. Then an
STM measurement of the local tunneling conductance $dI(V, x)/dV$ is taken over a 
predetermined $L\times L$ grid of points that cover a region of order 
600\AA$\times$600\AA.
Assuming that the tunneling conductance is
proportional to the underlying density of states of the $Cu\,O_2$ layer
\cite{MB01}, these measurements
give an STM map of the local tunneling density of states
 $N(x, \omega)$ with $\omega = eV$.  This map is then Fourier transformed, 
\begin{equation}
N(q,\omega) = \sum_{x_l\in(L\times L)} e^{-i q \cdot  x_l}
N(x_l,\omega)~,
\label{one}
\end{equation}
and the wave-vector power spectrum,
\begin{equation}
P(q, \omega) = \frac{|N(q, \omega)|^2}{L^2}~,
\label{two}
\end{equation}
determined.  Typically, the square root of the power spectrum, which is
proportional to the magnitude of $N(q, \omega)$, is plotted and we will follow
that practice as well. Here, we will discuss the structure of $P(q, \omega)$ and its
relationship to the quasi-particle spectrum and the structure factor
of the scattering potential.

For an isotropic $s$-wave superconductor with a circular normal state Fermi surface,
the ripples in $N(x, \omega)$ produced by a weak scattering center form a circular
pattern whose amplitude and wave length depend upon the bias voltage 
$\omega = eV$.
However, if the gap has $d_{x^2-y^2}$ symmetry, the ripples emanating from
a scattering center appear as a characteristic set of rays whose wave length
and amplitude vary with their angular direction and the size of the bias voltage
$\omega =  eV$\cite{BFS93,FB99}. Two examples of this are
shown in Fig.~1. 
It is the STM
measurements of these modulations in $N(x,\omega)$ that provide information on the wave vector and
frequency dependence of the gap. Indeed, in their FT-STM power spectrum study of BISCO,
Hoffman \textit{et.~al} \cite{Hof02} and McElroy \textit{et.~al} \cite{McE02}
 found frequency-dependent
structure in $P(q, \omega)$ which they argued were consistent with the Fermi surface dependence of $\Delta (k)$
as seen from ARPES measurements \cite{Din96}.
In this work \cite{Hof02,McE02}, these authors suggested that the FT-STM 
data could be analyzed in terms of a set of  frequency-dependent wave vectors 
${\bf q}_\alpha (\omega)$ which
connect the tips of the constant energy contours specified by the quasi-particle
dispersion relation
\begin{equation}
\omega = \sqrt{\epsilon^2_k + \Delta^2_k}.
\label{three}
\end{equation}
The contours of the solid regions shown in Figs.~1(b) and (d) are the
constant quasi-particle energy contours for a cylindrical Fermi surface 
$\epsilon_k=k^2/2m-\mu$ with a gap $\Delta_k=\Delta_0\cos{2\theta}$ for
$\omega=0.5\Delta_0$ and $1.1\Delta_0$, respectively.
Also shown in Fig.~1(b) are several of the ${\bf q}_\alpha(\omega)$ 
wave vectors introduced
in Ref.~\cite{McE02}.  The wave vector ${\bf q}^\prime (\omega) $ is an
additional wave vector that we will discuss.
The wave vectors
${\bf q}_1 (\omega)$ and  ${\bf q}_5 (\omega)$, 
along with their symmetry-related counter parts (not shown), are
associated with the structure of the ripples in $N(x,\omega)$ seen along the
$x$ and $y$ axes of Fig.~1(a).  
Likewise, the wave vectors ${\bf q}_3 (\omega)$, ${\bf q}^\prime (\omega)$, and ${\bf
q}_7 (\omega)$ determine the structure of the ripples along the 45$^\circ$ axes of
Fig.~1(a).
In conjunction with this experimental work, Wang and Lee \cite{WL02} recently 
reported numerical calculations for the case of a single impurity which clearly showed
the quasi-particle interference arising from ${\bf q}_1 (\omega)$ and 
${\bf q}_7 (\omega)$.
In addition, as $\omega$ is varied these calculations showed
that a rich, kaleidescope-like, structure appears in the wave vector power 
spectrum when it is folded back into the first Brillouin zone.

In a similar way, for $\omega=1.1\Delta_0$, the wave vectors $Q_\alpha$
shown in Fig.~1(d), determine the structure of $N(x,\omega)$ seen
in Fig.1~(c). Here the ripples along the $x$ and $y$-axes are associated with
$Q_2$ and $Q_3$ (and their symmetry related counterparts in the $y$ direction),
while those along the diagonal are associated
with $Q_1$ (and its counterparts). In practice, when $\omega\gtrsim\Delta_0$,
inelastic scatering leads to a damping of the ripples in $N(x,\omega)$,
making the structure in $|N(x,\omega)|$ associated with the $Q_\alpha$
wave vectors difficult to detect.

Now, the quasi-particle interference pattern shown in Figs.~1(a) and (c) are 
for a single impurity. For a particular surface region over which the
STM measurements are made, there will be an array of scatterers leading to
a complex overlap of ripples. Here we will discuss how
the Fourier transformed wave vector power spectrum $P(q,\omega)$ allows
one to disentangle the quasi-particle interference effects from
the static structure factors of the scatterers.
In Section II we show that for the case of a 
weak scattering potential, $P(q, \omega)$ factors into one piece
which contains information on the nesting properties of the Fermi 
surface times a piece which is proportional to the static 
structure factor of the scatterers. We also note that 
these measurements contain information on the one
electron self-energy. Various examples are analyzed to show 
the type of information that is in principle contained in the FT-STM data.  
In Section III, the case of a layered 2D superconductor is studied with results for both $s$-wave and $d$-wave gaps discussed.
Section IV contains our conclusions and 
Appendices A and B contain more details of the calculations.

\section{FT-STM Power Spectrum of a 2D Normal Metallic Layer}

To begin, we first consider the case of a normal metallic 2D layer.
Suppose it is exposed to a weak potential
\begin{equation}
V=\sum_{s} \int d^2x \, \delta\epsilon(x)\, \psi^\dagger_s (x)\, \psi_s (x).
\label{six}
\end{equation}
with $\delta\epsilon(x)$ an energy shift at $x$.  For a BISCO-like
system this local energy change $\delta\epsilon(x)$
could arise from secondary effects associated with disorder away
from the $CuO_2$ plane, the regular potential of the $BiO_2$ layer or possibly a
weak static stripe potential.  This interaction creates a ripple in the one-electron
Green's function leading to a modulation in the local tunneling density of
states.  For the case of a weak potential which we will focus on, a Born 
approximation is appropriate so that the single particle Green's function is
given by
\begin{equation}
G(x, x^\prime, \omega) = G_o \left(x-x^\prime, \omega\right) +
\int d^2{x^{\prime\prime}} \, G_o (x-x^{\prime\prime}, \omega)\, \delta \epsilon
(x^{\prime\prime})\, G_o (x^{\prime\prime}-x^\prime, \omega).
\label{seven} 
\end{equation}
Here, $G_o(x, \omega)$ is the Green's function of the unperturbed system.
Then the change in the single spin
tunneling density of states at position $x$ is given by
\begin{equation}
\delta N(x, \omega) = - \frac{1}{\pi}\ Im \int d^2{x^{\prime\prime}} 
\, G_o(x-x^{\prime\prime}, \omega)\, \delta\epsilon\, (x^{\prime\prime})
\, G_o (x^{\prime\prime} - x, \omega)
\label{eight}
\end{equation}
Taking the spatial Fourier transform of $\delta N(x, \omega)$, 
on the $L\times L$ grid of points $\{{\bf x}_l\}$ specified by the 
STM measurements, one finds for $q\ne 0$, that 
\begin{equation}
N(q, \omega) = \sum_{x_l\in(L\times L)} e^{-i {q} \cdot {x_l}} 
\delta N(x_l, \omega)
= - \frac{\delta\epsilon\, (q)}{\pi}\ Im\, \Lambda\, (q, \omega)
\label{nine}
\end{equation}
with
\begin{equation}
\Lambda\, (q, \omega) = \int d^2x\,
\, e^{i{q} \cdot {x}} 
\, G_o(x, \omega)\, G_o(-x, \omega)
\label{ten}
\end{equation}
and
\begin{equation}
\delta\epsilon(q) = \int \frac{d^2x}{a^2} \,
\delta\epsilon(x) e^{-i q \cdot {x}}~.
\label{eleven}
\end{equation}
Here $a$ is the lattice spacing of the $L\times L$ STM grid
and $q=2\pi(n_x,n_y)/La$ with $n_x$ and $n_y$ integers running
from $-L/2+1$ to $L/2$.
The wave-vector power spectrum of the local tunneling density of states,
Eq.~(\ref{two}), is therefore given by
\begin{equation}
P(q, \omega) = \frac{1}{N}\ \biggl|\frac{1}{\pi} Im\, \Lambda\, (q,
\omega)\biggr|^2\ |\delta\epsilon(q)|^2.
\label{twelve}
\end{equation} 
Here $N=L\times L$ is the number of sites in the sampled region. Thus, in
the weak scattering Born approximation, $P(q, \omega)$ separates into a
piece $|Im\, \Lambda\, (q, \omega)/\pi|^2$ which describes the
quasi-particle interference and a piece $|\delta\epsilon(q)|^2$ which is
the static structure factor, $S(q)$, of the scattering potential.


For the random impurity case, one could imagine making STM maps over a large
number of different $L \times L$ regions.  Then by averaging the structure
factor over these maps one would obtain 
\begin{equation} 
\frac{\langle|\delta\epsilon(q)|^2\rangle}{N} = n_i\delta\epsilon^2
\;\;\; (q \neq 0)
\label{avgstructf}
\end{equation} 
where
$n_i$ is the area impurity concentration 
and $\delta\epsilon$ is a site energy shift.  In this case, $P(q, \omega)$ would
simply be proportional to $n_i \delta \epsilon ^2 $ times the quasi-particle
interference factor from a single impurity.  However, this is not the way the
experiments are done.  Rather, a single STM map on a finite $L\times L$ grid of
points covering a specific region is measured.  In this case, $|\delta 
\epsilon(q)|^2$ versus $q$ exhibits fluctuations bluring the image of
$|Im \Lambda(q, \omega)/\pi|$, although one can still resolve
structure in false color 2D $(q_x,q_y)$ maps of $|N(q,\omega)|$. 
However, as discussed in 
Appendix A, by averaging the power spectrum over blocks of width 
$(\Delta q_x, \Delta q_y)$ about each $q$, the fluctuations can be reduced
if the impurities are randomly distributed.  
Naturally, this reduces the momentum resolution.  However, if the change in 
$q$ of the quasi-particle interference response is predominately along a given 
momentum direction, one can average over a region of $q$ values 
perpendicular to the direction of interest, reducing the fluctuations but 
maintaining the $q$ resolution in the direction of interest.  Here we will 
assume that a suitable average has been done and use the impurity structure 
factor given by Eq.~(\ref{avgstructf}). Appendix A contains a further
discussion of the effect of impurity induced fluctuations.

For the case in which the scattering occurs from a regular lattice such
as the $Bi$ lattice, one has
\begin{equation}
\left|\delta\epsilon (q)\right|^2 = 
\delta\epsilon^2 N^2 \delta_{{ q}, {G}_n} 
\label{fourteen}
\end{equation}
Here ${\bf G}_n$ is a reciprocal lattice vector of the $Bi$ lattice along
with the satellite wave vectors associated with the supermodulation of
the $BiO_2$ layer.
One could also have a ``random'' array of stripe domains with
\begin{equation}
\left\langle\left|\delta\epsilon (q)\right|^2\right\rangle \simeq N_i
\left\{\frac{\Gamma/\pi}{(q_x-Q_x)^2 + \Gamma^2} + 
\frac{\Gamma/\pi}{(q_y-Q_y)^2 + \Gamma^2}
+ (q_x \rightarrow -q_x, q_y \rightarrow -q_y)
\right\}\, \delta\epsilon^2
\label{fifteen}
\end{equation}  
Here $Q_x=2\pi/l_x$ with $l_x$ the stripe spacing, 
$2\pi/\Gamma$ is the characteristic size of a domain 
and $N_i$ is the average number of domains in an $L\times L$ region.
Here we have taken only the first  $Q_x$ harmonic. The form factor
associated with the charge distribution of the stripes suppresses
the response at higher multiples of $Q_x$. We will examine
the effect of an array of scattering centers in Section III.

Turning next to the quasi-particle interference response, we
begin by looking at $G_o(x, \omega)$ for a free 2D electron gas.  In this case, for
$\omega>0$,
\begin{equation}
G_o(x, \omega) = \int\, \frac{d^2k}{(2\pi)^2}
\ \frac{e^{i{k} \cdot {x}}}{\omega-\epsilon_k + i\delta}
= -i\, \pi N(0)\, H^{(1)}_0 \left(k(\omega)r\right),
\label{seventeen}
\end{equation}
with $\epsilon_k=k^2/2m-\mu$ and $mu=k_F^2/2m$.
Here $N(0)=m/2\pi$ is the single spin electron density of states for the 2D free
electron gas,
$H^{(1)}_0$ is the zeroth order Hankel function of the first kind, $r=|x|$ and
\begin{equation}
k(\omega) = k_F \sqrt{1+\frac{\omega}{\mu}}.
\label{eighteen}
\end{equation}
When $k(\omega)r$ is large
\begin{equation}
G_0(x, \omega) \sim - i\, N(0)\ \left(\frac{2\pi}{k(\omega)
r}\right)^{\frac{1}{2}}\ e^{i\left(k (\omega)r-\frac{\pi}{4}\right)}.
\label{nineteen}
\end{equation}
and the spatial modulation of $\delta N(x, \omega)$ which 
varies as the square of 
$G_o(x, \omega)$ is characterized by
a wave vector
\begin{equation}
q\, (\omega) = 2k_F \sqrt{1+ \frac{\omega}{\mu}}.
\label{twenty}
\end{equation}

The quasi-particle interference response function $\Lambda (q, \omega)$
for the 2D electron gas is calculated in Appendix B.  The result of
this calculation shows \cite{Kivup} that 
\begin{equation}
\frac{1}{\pi}\, Im \Lambda\, (q, \omega) = \Biggl\{\begin{array}{lr}
\frac{8\pi N^2(0)}{q\sqrt{q^2 - 4k^2(\omega)}}
& {\rm for}\  q>2k(\omega)\\
0 & {\rm for}\ q<2k(\omega)\end{array}
\label{twentyone}
\end{equation} 
Thus, the wave-vector power spectrum for the 2D free electron gas has a
cusp at $q$ equals $2k(\omega)$. $Im \Lambda(q,\omega)$ vanishes for
$q<2k(\omega)$ and diverges as $(q-2k(\omega))^{-1/2}$ as $q$
approaches $2k(\omega)$ from larger values. As noted in the appendix,
$Re\Lambda(q,\omega)$ has a similar cusp as $q$ approaches $2k(\omega)$
from below. Basically, there is just a shift of phase of $\pi/2$ in
$\Lambda(q,\omega)$ when $q$ passes trough $2k(\omega)$. 
If the impurities are dilute, but the scattering from a given impurity 
is strong, one still has
$P(q, \omega)$ proportional to the impurity concentration.
However, in this case 
\cite{thnkKivelson}, because of the phase of the t-matrix one 
will have singularities on both sides of $2 k(\omega)$. 


The experimental FT-STM data has been reported as the square root of the power
spectrum or ``the magnitude'' $|N(q, \omega)|$ of the Fourier transform of the STM
measurement of the local conductance map.  Here we will follow this convention.
For the case of weak Born scattering, the magnitude 
of $N(q, \omega)$ is proportional to $|Im\, \Lambda\, (q, \omega)/\pi|$.
For the 2D electron gas, in Fig.~2, we have plotted $|N(q, \omega)|$ normalized to
$\sqrt{n_i}N(0)|\delta\epsilon|/\mu$ versus $q$ for various values of $\omega/\mu$.  Here, with
$q$ in units of $2k_F$ one has
\begin{equation}
\left|\bar N (q, \omega)\right| = \frac{|N(q, \omega)|}{\sqrt{n_i}
\ N(0)\left(\frac{|\delta\epsilon|}{\mu}\right)} = \frac{1}{2q}
\ \frac{1}{\sqrt{q^2-\left(1+\frac{\omega}{\mu}\right)}}
\label{twentytwo}
\end{equation}
for $(1+\omega/\mu)<q/2k_F$.  As shown in Fig.~2, for the weak scattering case
$|\bar N(q, \omega)|$ has a 
one-sided square root singularity at $q/2k_F = (1+\omega/\mu)^{1/2}$. This singularity
is cut off and the response peak 
varies as $(\ell k_F)^{1/2}$ when the quasi-particle
mean-free path is taken into account.  In this case
\begin{equation}
\left|\bar N(q, \omega)\right| = \frac{1}{2\pi\, q}\ \int\limits^{q^2-1}_{-1}
d\epsilon
\ \frac{\Gamma}{\left(\frac{\omega}{\mu}-\epsilon\right)^2
+ \Gamma^2}\ \frac{1}{\sqrt{q^2-(1+ \epsilon)}}
\label{twentythree}
\end{equation} 
with $\Gamma=(\ell k_F)^{-1}$.
Plots of $|\bar N(q, \omega)|$ versus $q$ for $\omega=0$ and several different
values of $\Gamma$ are shown in Fig.~3.

Similarly, for an anisotropic system
with
\begin{equation}
\epsilon_k = \frac{k^2_x}{2m_x} + \frac{k_y^2}{2m_y} - \mu
\label{twentyfour}
\end{equation}
one finds that
\begin{equation}
\frac{1}{\pi}\ Im\, \Lambda\, \left(q_x, q_y, \omega\right) = 
\left\{ \begin{array}{lccr}
\frac{2}{\pi\gamma}  m^2_x 
& \frac{1}{\sqrt{q^2_x + \gamma^2 q^2_y}} 
& \frac{1}{\sqrt{q^2_x + \gamma^2 q^2_y-8m_x (\omega+ \mu)}}
& {\rm for}\ q^2_x + \gamma^2 q^2_y > 8m_x (\mu + \omega)\\
0 &&& {\rm for}\ q^2_x + \gamma^2 q^2_y < 8m_x (\mu + \omega) \end{array} \right.
\label{twentyfive}
\end{equation}
with $\gamma^2 = m_x/m_y$.  In this case the cusp in the power spectrum
follows a locus determined by 
\begin{equation}
q^2_x + \gamma^2q^2_y = 8m_x (\mu + \omega)
\label{twentysix}
\end{equation}
which reflects the elliptical Fermi surface. In Fig.~4, we have plotted $|N(q,
\omega)|$ for $\omega=0$ and $\gamma=3$ normalized to $\sqrt{n_i}(m_x/2\pi)
|\delta\epsilon|/\mu$, which gives
\begin{equation}
\left|\bar N(q_x, q_y, \omega)\right| = \frac{1}{2\gamma}
\ \frac{1}{\sqrt{q^2_x + \gamma^2 q^2_y}}
\ \frac{1}{\sqrt{q^2_x + \gamma^2 q^2_y - \left(1+\frac{\omega}{\mu}\right)}}
\label{twentyseven}
\end{equation}
with $q_x$ and $q_y$ measured in units of $2k_{F}^x=2(2m_x\mu)^{1/2}$.  
Here, we see that $|\bar N(q, 0)|$
vanishes inside the ellipse Eq.~(\ref{twentysix}) and has a square root divergence
as $q$ approaches the ellipse.
There is a reduction $\gamma=(m_y/m_x)^{1/2}$ in the strength of the cusp
along the $q_y$ direction relative to the $q_x$ direction that
reflects the fact that the joint density of
states which enters $\Lambda\, (q, \omega)$ depends on the curvature of the Fermi surface.

Turning now to the case of a tight-binding band, we consider first the simple
nearest-neighbor hopping band for a square lattice with a unit lattice spacing
\begin{equation} 
\epsilon_k = - 2t\, \left(\cos k_x + \cos k_y\right) - \mu.
\label{thirty}
\end{equation}
The Fermi surface is shown in Fig.~5 for $\mu=-t$, corresponding to a small filling
$\langle n \rangle \approx 0.31$ which we have chosen to illustrate what happens when $Im \Lambda(q, w)$ is folded back into the first Brillouin zone. In the following tight-binding bandstructure 
calculations,
we will assume that $N(x, \omega)$ is measured on a grid of points corresponding to
the sites of the lattice.  In this case, 
\begin{equation}
\Lambda(q, \omega) = \frac{1}{N}\, \sum_k\, G_0(k+q,\omega)\, G_0(k,\omega)
\label{thirtyone}
\end{equation}
with
\begin{equation}
G_0(k) = \frac{1}{\omega-\epsilon_k + i \delta \, sgn (\omega)}
\label{thirtyonea}
\end{equation}
Here, $N=L\times L$ is the number of lattice sites and $k$ and $q$ are
defined in the first Brillouin zone with components running from $-\pi$ to $\pi$.
We have set the lattice spacing $a=1$. This
choice of the grid simplifies the calculations but has the consequence that all
results are folded back into the first Brillouin zone. It is this down folding that
leads to the kaleidoscopic patterns in the numerical results shown in Ref.~\cite{WL02}. 
In the
experimental FT-STM measurements \cite{Hof02,McE02}, a smaller
grid spacing was used
leading to a $q$-space power spectrum which looks more like the 
extended zone picture for the $q$-vectors of interest.

Carrying out the momentum sum in Eq.~(\ref{thirtyone}), we find the results for
$|Im\Lambda \, (q,\omega)/\pi|$
shown in Fig.~6. Here in Fig.~6(a), $q$ varies along the diagonal a-cut shown in
Fig.~5 with $q_x=q_y$. In this case, $|\bar N(q, \omega)|$ exhibits a similar cusp to
that of the free electron system when $|q_x| \geq 2k_F^{xy}= 2 \cos^{-1} (-(\mu + \omega)/4t)$.
This same type of behavior is shown as the dashed curve for the b-cut with 
$q_y=0$ in
Fig.~6(b), where we have displaced the numerical results by $\pm 2\pi$ corresponding to
an extended zone scheme.  Here, the cusp occurs for $|q_x| \geq 2k_F^x=
2\cos^{-1} (-(2t+\mu+\omega)/2t)$ which is greater 
than $\pi$ for the b-cut of Fig.~5.
In practice, the numerical data is obtained in the ``reduced'' $(-\pi, \pi)$
zone so that the $q_y=0$ b-cut appears as the solid curve shown in Fig.~6(b).
As $\omega$ is increased, the
characteristic $q$ values can move across the boundary of the first zone and be
mapped back via a reciprocal lattice vector.  This can then lead to a situation in
which the square root singularity is approached from smaller $q$-values with $|\bar
N(q, \omega)|$ vanishing when $q$ exceeds a critical value as shown by the solid
curve in Fig.~6(b). As noted, it is
this folding of the FT-STM power spectrum into the reduced zone that 
leads to the kaleidoscopic 2D $(q_x,q_y)$ patterns 
for different $\omega$ values which have been reported \cite{WL02}.

Finally, consider the case of a tight-binding band, 
like that found for BISCO. Here, one has a next-near-neighbor 
hopping $t^\prime$ so that
\begin{equation}
\epsilon_k=-2t(\cos{k_x}+\cos{k_y})-4t^\prime\cos{k_x}\cos{k_y}-\mu~.
\label{tb}
\end{equation}
The Fermi surface for $t^\prime/t = -0.3$ and $\mu/t = -1.0$ is shown in
Fig.~7. Results of a numerical calculation for $Im \, \Lambda
\, (q, \omega)/\pi$ for this case are plotted in
Figs.~8(a) and 8(b) for $q_y=0$ and $q_x=q_y$, respectively. 
For $q_y=0$, the nesting
vector 3 is shown in Fig.~7. If we consider the 
closed Fermi surface around $(\pi,\pi)$, one can see that the two 
points of this closed Fermi surface that are connected
by the wave vector labeled 3, have 
$\Delta q$ greater than $\pi$. In fact, $\Delta
q/\pi \simeq 1.76$, so that when this is folded back into the first Brillouin zone, the
cusp occurs at $\Delta q/\pi \simeq - .24$. Also, as we saw previously in Fig.~6(a) the cusp
rises from smaller values of $q$ in the reduced zone although in an extended zone
scheme it would be approached from larger $q$ values just as in the free electron
case.  The peaks in the response for $q_x=q_y$ shown in Fig.~8(b) arise from the nesting
vectors labeled 1 and 2 in Fig.~7.

To conclude this Section on the normal state, 
we consider an electron-phonon system with a self-energy
$\Sigma(\omega)$ which depends only upon the frequency.  In this case,
the momentum that enters the Hankel function giving $G(x, \omega)$ is 
\begin{equation}
k(\omega) \cong k_F + \frac{(\omega - \Sigma (\omega)}{v_F}.
\label{thirtytwo}
\end{equation}
The real part of the self-energy leads to a shift in the wave length
of the modulations and the imaginary part of $\Sigma$ leads to their
exponential decay on a scale $\ell = v_F/(-2\Sigma_2 (\omega))$.  This
behavior is reflected in a shift in the position and a rounding of the
cusp in the power
spectrum $P(q, \omega)$. For values of $\omega$ which are small
compared to a typical phonon frequency
\begin{equation}
k(\omega) \simeq k_F + \frac{(1+\lambda)\omega}{v_F}
\label{thirtythree}
\end{equation}
with $\lambda = 2N(0) |g|^2/\Omega_0$ the dimensionless electron phonon
interaction strength.  Here $g$ is the effective electron-phonon
coupling and $\Omega_0$ a typical phonon energy. As $\omega$ increases $2k_F(\omega)$ will reflect the detailed
dependence of both the real and imaginary parts of $\Sigma(\omega)$.

\section{The FT-STM Power Spectrum of a Superconductor}

Next consider the case of a superconductor.  Here, for an on-site energy
perturbation, we have \cite{BFS93}
\begin{equation}
\delta N(x, \omega) = - \frac{1}{\pi}\ Im\, \int d^2{x^\prime} 
\left(G \left(x-x^\prime, \omega\right)G\left(x^\prime-x, \omega\right)
- F\left(x-x^\prime, \omega\right) F\left(x^\prime-x, \omega\right)\right)
\delta\epsilon\left(x^\prime\right)
\label{thirtyfour}
\end{equation}
with $G(x,\omega)$ the usual single particle propagator
\begin{equation}
G(x, \omega) = \int \frac{d^2k}{(2\pi)^2} \, e^{i k\cdot x}
\, \frac{\omega + \epsilon_k}{\omega^2 - \epsilon^2_k - \Delta^2_k + i\delta}
\label{thirtyfive}
\end{equation}
and $F(x,\omega)$ the anomalous Gor'kov propagator
\begin{equation}
F(x, \omega) = \int \frac{d^2k}{(2\pi)^2} \, e^{i k\cdot x}
\, \frac{\Delta_k}{\omega^2 - \epsilon^2_k - \Delta^2_k + i\delta}.
\label{thirtysix}
\end{equation}
In this case \cite{ref2}
\begin{equation}
\Lambda\, (q, \omega) = \int \frac{d^2k}{(2\pi)^2}
 \biggl(G(k+q, \omega)\, G(k, \omega)
- F\, (k + q, \omega) F\, (k, \omega)\biggr)
\label{thirtysixa}
\end{equation}
For a 2D $s$-wave superconductor with 
$\Delta_k = \Delta$ and $\epsilon_k=k^2/2m-\mu$, these Green's functions
have characteristic wave vectors
\begin{equation}
k_{\pm}(\omega) = k_F \left[1 \pm 
\frac{\sqrt{\omega^2-\Delta^2}}{\mu}\right]^{\frac{1}{2}}
\label{thirtyseven}
\end{equation}
The wave length of the ripples in $\delta N(x, \omega)$ are set by
$2k_{\pm}(\omega)$. For $\omega < \Delta$,
the ripples exponentially decay.  If the impurity perturbation involved a change in
the size of the gap, there would be a quasi-particle interference contribution
involving $k_+ (\omega) - k_- (\omega) = 2k_F \sqrt{\omega^2-\Delta^2}/\mu$. The
ripples associated with this contribution vary on a length scale set by $v_F/\Delta$
which is of order the coherence length rather than $k_F^{-1}$.  
It is this type of spatial  oscillation that is responsible 
for the Tomasch oscillations \cite{Tom65}. While Tomasch ripples at
$Q_T(\omega)=k_+(\omega)-k_{-}(\omega)$ are not present for the case
of a charge impurity in an isotropic $s$-wave superconductor, they can appear
for a $d$-wave superconductor as discussed below.

As shown in Appendix B, it is straightforward to evaluate $ Im\, \Lambda
\, (q, \omega)$ for an $s$-wave BCS superconductor and one finds that 
\begin{equation}
\frac{1}{\pi}\, Im\, \Lambda\, (q, \omega) = 4\pi\, N^2 (0)
\ \left(\frac{\omega-\sqrt{\omega^2-\Delta^2}}{\sqrt{\omega^2-\Delta^2}}\right)
\ \frac{1}{q}\ \frac{1}{\sqrt{q^2-4k^2_-(\omega)}}
\label{thirtyeight}
\end{equation}
for $2k_{-} (\omega) < q < 2k_{+}(\omega)$ and
\begin{eqnarray}
\frac{1}{\pi}\, Im\, \Lambda\, (q, \omega) & = & 4\pi\, N^2 (0)
\left(\frac{\omega-\sqrt{\omega^2-\Delta^2}}{\sqrt{\omega^2-\Delta^2}}\right)
\ \frac{1}{q}\ \frac{1}{\sqrt{q^2-4k^2_-(\omega)}}\nonumber\\
& + & 4\pi\, N^2(0) \left(\frac{\omega +
\sqrt{\omega^2-\Delta^2}}{\sqrt{\omega^2-\Delta^2}}\right)
\ \frac{1}{q}\ \frac{1}{\sqrt{q^2-4k^2_+ (\omega)}}
\label{thirtynine}
\end{eqnarray}
for $2k_+(\omega) < q$.  
More generally, for an electron-phonon system the
frequency-dependent complex gap $\Delta(\omega)$ would replace $\Delta$. 
Normalizing by $\sqrt{n_i}N(0)\, |\delta\epsilon|/\mu$, as previously done for the free
electron case, we have plotted $|\bar N(q, \omega)|$ versus $q$
in Fig.~9 for a BCS $s$-wave superconductor with $\Delta/\mu = 0.1$.  The coherence
factors in Eq.~(\ref{thirtyeight}) and (\ref{thirtynine})
 give more weight to the $2k_+(\omega)$ cusp for
positive values of $\omega$ associated with a bias voltage that probes the empty
states, Fig.~9(a),  while the $q=2k_-(\omega)$ cusp is enhanced when the bias is reversed
as shown in Fig.~9(b).

We turn next to the case of a $d$-wave superconductor with 
\begin{equation}
\Delta_k = \Delta_0 (\cos k_x - \cos k_y)/2
\label{fortyone}
\end{equation}
and a bandstructure given by Eq.~(\ref{tb}) 
with $t^\prime/t = -0.3$ and $\mu = - 1.0$.  
The Fermi surface for these parameters is shown as a dashed line 
in Fig. 10(a). The contours of the solid regions correspond to the loci of
points where $\omega=\sqrt{\epsilon_k^2+\Delta_k^2}$ for $\omega=0.5\Delta_0$
in Fig. 10(a) and $\omega=1.1\Delta_0$ in Fig.10(b). 
Results for $|Im\, \Lambda\, (q, \omega)/\pi|$
at various values of $\omega$ less than $\Delta_0$ are shown in Fig.~11(a) 
for $q_x=q_y$ and Fig.~11(b) for $q_y=0$.  For the diagonal cut with $q_x=q_y$, there is a response at the wave vector
${\bf q}_7(\omega)$ shown in Fig.~10(a) which connects the end points of the 
$\omega^2 = \epsilon^2_k + \Delta^2_k$ contour.  This peak is similar to the peak we
have seen in the case of the ellipse discussed in the previous section.  There are two
identical contributions coming from the two contours on 
opposite sides of the Fermi
surface.  As the bias voltage $eV=\omega$ increases, one sees 
that $q_7(\omega)$ peak moves to larger values of momentum.
This reflects the increase in magnitude of the wavevector $q_7(\omega)$
as $\omega$ is increased and can provide information on the
$k$-dependance of the Fermi surface and gap \cite{Hof02,McE02}.
In addition, there is a contribution coming from ${\bf q}_3(\omega)$ which
connects the tips of two opposite contours which increases more slowly 
with $\omega$.  Finally, there is a response 
associated with ${\bf q}^\prime$ shown in Fig.~10(a).  This is the response
due to the wave vector
labeled 1 in Fig.~7 and shown in the diagonal $q_x=q_y$ response of the normal metal
with this same bandstructure in Fig.~8(b).
Approximate analytic results for the $d$-wave case are given 
in Appendix B, Eqs.~(\ref{athirteen}) and (\ref{asixteen}).

Similarly in Fig.~11(b), for $q_y=0$ 
one finds structure associated with the wave
vectors ${\bf q}_1$ and ${\bf q}_5$ shown in Fig.~10(a).  
As discussed by McElroy \textit{et. al}, \cite{McE02}
this structure arises from a peak in the joint density of 
states associated with the overlap of the ends 
of two opposite $\omega^2 = \epsilon^2_k + \Delta^2_k$ contours.  
Just as for the case of the elliptical Fermi surface 
previously discussed, the strength of the cusp depends upon the 
curvature of the dispersion. The increase of $q_1(\omega)$ with increasing 
$\omega$ again provides information on the Fermi surface and $\Delta_k$.
The peak at $q_5(\omega)$ initially increases with $\omega$ and
would continue to increase in an extended zone but here, 
for $\omega\gtrsim0.25$, it is reflected back into the first Brillouin zone.
Finally, there is a weak cusp at low momentum $Q_T(\omega)$ associated
with the Tomasch interference process.

Note that the response seen in 
$|\bar N(q, \omega)|$ associated with quasi-particle interference is actually
characterized by a continuous curving cusp in the $(q_x, q_y)$ plane
whose intensity is related to the joint
density of the $k$ and $k+q$ quasi-particle states.
When we take into account all four quadrants, the structure 
in the power spectrum $P(q,\omega)$ in the first 
Brillonin zone becomes much richer. Not only are there
cusps associated with scattering processes confined to similar 
contour regions in quadrants 2 and 4, which give rise 
to a cusp whose major axis is perpendicular to that due to quadrants 
1 and 3, there are cusps from scattering processes between the four
nodal regions.  
In addition, there is the response associated with the nodal wave vector
${\bf q}^\prime$.  
In order to see this more clearly, 
we have calculated $\Lambda\, (q, \omega)$ for the case in
which $t^\prime=0$.  
In Fig.~12 we compare the results for $|Im\, \Lambda\, (q,\omega)/\pi|$ 
calculated with $t^\prime=0$ [Fig.~12(a)] and with $t^\prime/t=-0.3$
[Fig.~12(b)].  In both cases, we clearly see the contributions of ${\bf q}_3$ 
and ${\bf q}_7$.  In addition, a ${\bf q}^\prime$ 
contribution from the nodal region is also visible.  The ${\bf q}_3$
contribution is weaker for the case in which $t^\prime=0$.  
In this case we also find that $|{\bf q}_3| < |{\bf q}^\prime|$ 
as one can see from the insets in the figures. 

When $\omega$ is greater than $\Delta_0$, structure in $|N(q,\omega)|$
is associated with the $Q_\alpha$ vectors shown in Fig.~10(b), for 
$\omega=1.1\Delta_0$. In this case, for $q_x=q_y$, $|N(q,\omega)|$
exhibits structure at $q_x=Q_1$, $Q_2$, and $Q_3$, see Fig.~13(a).
The weak structure at large momentum transfer $Q_1$ is similar to the 
structure for the normal state labeled "1" in Fig.8, since the gap
vanishes along the diagonal. The structures at $Q_2$ and $Q_3$
correspond to interference processes between the particle like
and the hole like BCS quasi particles. Just as for
the $s$-wave case, for $\omega>0$, the coherence factor gives
more weight to the $Q_2$ process. Results for $|N(q,\omega)|$
versus $q_x$ for $q_y=0$ are shown in Fig.~13(b). In this case,
structure appears associated with $q_x=Q_4$ and $Q_5$ with the coherence
factors leading to a large response at $Q_4$ for positive $\omega$.
As noted earlier, in the cuprate superconductors 
lifetime effects suppress this $\omega>\Delta_0$ structure,
making it difficult to see in the experimental studies.

The structure in the quasi-particle interference response
can also be seen in the intensity map plots of 
$|Im\, \Lambda\, (q,\omega)/\pi|$ over the $(q_x,q_y)$ plane. 
One such map for $\omega=0.5\Delta_0$ is shown in Fig.~14. Here one sees
that the response is characterized by continuous curving intensity cusps
in the $(q_x,q_y)$ plane. Like the case of the elliptical Fermi surface, 
the intensity can have significant variations along the cusps due to the
curvature of the quasi-particles dispersion relation. Going out from the 
origin along $q_x=q_y$ as in Fig.~12(b), one first sees a bright (high intensity
region) cusp associated with $q_7$. At large momentum values
along this same 45$^\circ$ line one sees a narrow bright line
associated with $q^\prime$ scattering processes which connect 
the outter edges of the 
$\omega = \sqrt{\epsilon^2_k + \Delta^2_k}$ envelopes.
Finally, the brighter, curved region of intensity near the $(\pi,\pi)$
corner arises from $q_3$ interference processes associated
with the inner boundaries of these contours.
The bright regions near $(q_x/\pi=0.85,q_y/\pi=0.3)$ and
$(q_x/\pi=0.3,q_y/\pi=0.85)$ arise from interference
effects associated with $q_2$ and $q_6$ of Fig.~10(a).

Finally, we consider the response of a $d$-wave superconductor
when there is an ordered stripe array of weak scattering centers.
In Fig.~15 we show the ripples produced in $N(x,\omega)$ when
the scattering centers form a stripe-like structure oriented along
the $y$-axis with a spacing of four lattice sites. For the parameters
we have chosen, $q_5(\omega)\simeq \pi/2$ for $\omega=0.5\Delta_0$
with $\Delta_0$ the maximum value of the $d$-wave gap. As seen in
Fig.~15, the ordered array of stripe scattering centers produces
an oriented set of ripples. These give rise to the structure
in $|N(q,\omega)|$ shown in Fig.~16 for $q\simeq \pi/2$. Here
we have assumed that there is quasi-particle scattering from a
random array of impurities as well as the stripes. These
contributions add incoherently so that
\begin{equation}
\left\langle\left|\delta\epsilon (q)\right|^2\right\rangle \sim
\left[ 1+ A\frac{\Gamma/\pi}{(q_x-Q_x)^2+\Gamma^2}\right]~.
\end{equation}
Here $Q_x=\pi/2$ and we have taken $\Gamma=0.01$ and 
$A=0.5$ in Fig.~16 to illustrate
the possible interplay of the scattering from the random impurities and 
the stripes. The quasi-particle interference peak associated with $q_1$
moves to lower values of $q_x$ as $\omega$ increases while
the response associated with the ordered array of scattering
centers remains fixed at $q_x\simeq 0.5\pi$ and only its amplitude
changes with $\omega$. For larger values of $A$, the stripes would be
the dominant feature while for smaller values of $A$ scattering from
the random impurities would dominate the power spectrum.

\section{Conclusions}

Here we have discussed some detailed examples which illustrate what can be learned from FT-STM
studies of layered materials.  For the case of a weak potential perturbation we have
seen that the wave vector power spectrum of the local tunneling density of states,
$P(q, \omega)$, contains information on the quasi-particle spectrum and the structure
factor of the scatterers. This is also the case for dilute impurities even if they act as strong scattering centers and $\delta \epsilon$ is replaced by a t-matrix.  For a normal
metal, one can determine information about
the nesting properties of the Fermi surface from the loci of the ${\bf q}$-space 
cusps for
$\omega\to 0$.  In addition, the dressed Fermi velocity can be obtained from the
$\omega$ dependence of the position of the cusp.  In fact, from the $\omega$ 
dependence of the cusp position and its rounding, one can, in principle, obtain
information on the real and imaginary parts of the electron self-energy.

In the superconducting state, one can also obtain information on both the wave
vector and frequency dependence of the gap $\Delta (k, \omega)$.  Again, just as in
the case of a normal metal, the signature of these quasi-particle interference effects
are continuous ``arcs'' of cusps in ${\bf q}$-space.  However, 
the intensity variation along the arcs
may be large due to changes in the effective density of states associated with a
given momentum transfer ${\bf q}$,
making the intensity appear more like spots.
The rapid change in intensity is due to the small
parameter $\Delta_0/t$.

In addition to these effects, one can also obtain information on the static structure
factor of the scattering potential $\langle |\delta\epsilon_q|^2\rangle$.
As discussed in Appendix A, for the case of a local potential, averaging over a $\Delta q_x$ by $\Delta q_y$ block of $q$ values about a given $q$ leads to $\langle |\delta\epsilon_q|^2\rangle/N \simeq n_i \delta\epsilon^2$.  However, if the scattering potential has
long-range order, such as is the case for the $BiO_2$ over layer in the BISCO studies,
one should see a $q$-dependent response in the wave-vector power spectrum of
the local tunneling density of states for ${\bf q}$ values equal to the 
reciprocal lattice
vectors of the $BiO_2$ layer.  This response will be particularly
strong for $\omega\simeq\Delta_0$.
Similarly, it should be possible to see evidence of
pinned stripes in the structure factor if they are present \cite{How02}. 

As discussed in the introduction, we have focused on the case
of weak scattering.  As noted, in this limit, where the Born approximation is
adequate, one has a simple separation of the response into the quasi-particle
interference effects which are contained in 
$\Lambda (q, \omega)$ and the structure
factor of the scatterers which is contained in $\langle |\delta
\epsilon_q|^2\rangle$. Of course, many-body interactions will give rise to 
additional effects.  
There will be screening, changing $\delta\epsilon_q$ to
$\delta\epsilon_q/\epsilon(q,0)$ 
where $\epsilon (q, 0)$ is the zero frequency dielectric constant \cite{Kivup}.  
Furthermore, there
will be vertex corrections so that for example, for a strongly-interacting normal system one will have
\begin{equation}
\Lambda (q, \omega) = \int \frac{d^2k}{(2\pi)^2}\ \Gamma\, (k, q, \omega)
\, G(k+q, \omega)\, G(k, \omega)~,
\label{fortythree}
\end{equation}
with $\Gamma(k,q,\omega)$ the elastic vertex for momentum 
transfer $q$ and zero energy transfer for an incoming particle with
momentum $k$ and energy $\omega$.
Similar vertex corrections will occur in the superconducting state.
In addition, the quasi-particle
dispersion relation can be altered by an interaction which breaks the 
translational symmetry \cite{PDDH02} 
leading to a different $\omega$ dependence of 
$|N(q,\omega)|$. 

Finally, there is the form factor of the tunneling probe.  Here we have
neglected the momentum dependence of the tunneling matrix element and simply
assumed that the conductance map was proportional to the local tunneling
density of states $N(x,\omega)$.  However, if this is not the case, a tunneling
matrix element form factor
\begin{equation}
T(k) = \sum_l e^{i k \cdot l} T(l)
\end{equation}
will enter so that for a superconductor
\begin{equation}
\Lambda(q, \omega) =\frac{1}{N} \sum_k T^*(k+q)T(k)
\left[ G(k+q, \omega) G(k, \omega) - F(k+q, \omega) F(k, \omega) \right]
\end{equation}
For example, a tunneling form factor 
\begin{equation}
T(k) = ( \cos k_x - \cos k_y)
\end{equation}
has been suggested for the case of an STM tip on a cuprate superconductor 
\cite{MB01}. However. because the average tunnelling density of states in 
these STM experiments appears to vary linearly with $V$ at low voltages, 
we have modeled the tunnelling as a direct process and neglected its
momentum dependence.

\appendix

\section{Random Impurity Structure Factor}
\newcommand\deq{\ensuremath{\delta\epsilon(q)}}
\newcommand\dqx{\ensuremath{\Delta q_x}}
\newcommand\dqy{\ensuremath{\Delta q_y}}
\newcommand\magnitude[1]{\ensuremath{\left| #1 \right|}}

Suppose we have an $L \times L$ section of a lattice with a concentration 
of $n_i$ impurities per unit area.  Assume that if there is an impurity at
site $l$ it has a potential $\delta \epsilon(l)=1$, while if there is
no impurity at site $l$, $\delta \epsilon(l) =0$.  For a
given random configuration of $N_i$ impurities on $L \times L= N$
sites, corresponding to an area impurity concentration $n_i = N_i/N$, we have
\begin{equation}
\deq = \sum_l e^{i q \cdot l} \delta \epsilon(l). 
\label{eqn:foo}
\end{equation}
For this configuration of impurities we can define a power
spectrum
with 
\begin{equation}
P(q) = \frac{\magnitude{\deq}^2}{N}
\label{eqn:powspec}
\end{equation}
There is of course a peak for $q=0$, where
\begin{equation}
P(0) = \frac{N_i^2}{N}.
\end{equation}
However for other values of $q$, $P(q)$ fluctuates about $n_i$.
If one were to average $P(q)$ over many realizations of independent
impurity configurations \footnote{
For a single impurity configuration
\[
\frac{1}{N} \sum_q P(q) =n_i 
\]
and if we drop the $q=0$ term then
\[
\frac{1}{N} \sum_{q \ne 0} P(q) =n_i(1-n_i) 
\]
},
\begin{equation}
\left\langle P(q) \right\rangle = n_i \ \mbox{for} \ q \ne 0.
\end{equation}

However, if we have an $L \times L$ lattice with a fixed configuration of
impurities, the power spectrum $P(q)$, given by
Eq.~(\ref{eqn:powspec}), will exhibit fluctuations.  If
we average $P(q)$ over blocks of $q$ of width $(\Delta q_x, \Delta
q_y)$ about each $q$ we can reduce these fluctuations.  Naturally at
the same time, the momentum resolution will be reduced.
Define a block smoothed $P(q)$ average as follows:
\begin{equation}
\bar{P}(q) = \frac{1}{N(\Delta q_x, \Delta q_y)} 
\sideset{}{'}\sum_{\dqx, \dqy} P(q),
\label{eqn:powspecsmooth}
\end{equation}
here the prime on the sum indicates that $P(0)$ is replaced
with $n_i$ so that the $q=0$ peak is not broadened from the smoothing operation.
The sum in Eq.~(\ref{eqn:powspecsmooth}) is over a set of q-points $(- \dqx /2,  \dqx/2)$ and $(- \dqy
/2, \dqy /2)$ about each $q$. $N(\dqx, \dqy)$ is the number
of sites in the $\dqx$ by $\dqy$ block. We expect that the RMS
fluctuations of $\bar{P}(q)$ will decrease inversely as the square
root of the number of $q$-values in the $\dqx$ by $\dqy$ block.

For example, consider a lattice with $L=500$ and one configuration of random
impurities with $n_i=0.01.$  The power spectrum $P(q)$, given by
Eq.~(\ref{eqn:powspec}), versus $q_x$ for $q_y=0$ is shown in Fig.~17(a).
Figures 17(b) and 17(c) show similar plots for the smoothed power spectrum
$\bar{P}(q)$, given by Eq.~(\ref{eqn:powspecsmooth}), for square blocks
with $\dqx = \dqy = 2\pi (2 d + 1) / L$ for $d=5$ and $d=10$
respectively.  Figure 17(d) shows $\bar{P}(q)$ for the case in which
only $\dqy = 2\pi (2 d +1) / L$ is averaged with $d=10$.  The solid line in each of these figures is $n_i=0.01$.

It is clear from the results shown in Fig.~17, that one can
reduce the fluctuations in the power spectrum by averaging it over a
region of q-space surrounding a given q-point.  This is of course just
complementary to taking the original $L \times L$ spatial lattice,
breaking it up into $L_B \times L_B$ blocks with 
$L_B= 2\pi/ \Delta q$, constructing the appropriate Fourier transform
for each block and then averaging over the blocks.  The more blocks
one has, the small the RMS fluctuations of the power spectrum of the
Fourier transform. Of course, dividing the $L \times L$ system up into
more blocks leads to a corresponding decrease in the momentum
resolution.  The momentum smoothing operation,
Eq.~(\ref{eqn:powspecsmooth}), is just another way of blocking.

As a further test of these ideas, we calculate the RMS deviation of
$\bar{P}(q)$ from $n_i$ versus the momentum block size $2d+1$
\begin{equation}
D = \sqrt{\frac{1}{N}\sum_{q \ne 0} \left(\bar{P}(q) - 
\frac{1}{N}\sum_{q \ne 0} \bar{P}(q)\right)^2}.
\label{eqn:deviation}
\end{equation}
In Fig.~18 we show that $D$ varies as $(2d+1)^{-1}$ for large $d$ for
several different concentrations of impurities.  The straight lines on
the plot represent the asymptotic behavior in which $D \sim \alpha n_i /
(2 d +1)$ with $\alpha$ of order one.

Finally, we model the behavior of the magnitude of the wave-vector
Fourier transform of the local tunneling density of states by 
\begin{equation}
\magnitude{N(q)} = \sqrt{P(q)} \times \left\{ \begin{array}{ll}
\frac{1}{\sqrt{q - q^\star}} & \mbox{if $q> q^\star = \pi/4$} \\
0 & \mbox{otherwise}.
\end{array} \right.
\label{eqn:nq}
\end{equation}
The results for $L=500$ and $n_i=0.01$ are shown in Figs.~19(b)-(c) for the
$\mbox{BlockAvg}\left[\magnitude{N(q_x, q_y=0)}\right]$, averaged over a $\dqx
\times \dqy$ square with $d=5$ and $d=10$ respectively.  Figure 19(a) shows the
unaveraged result.  Similarly Fig.~19(d) shows the
$\mbox{LineAvg}\left[\magnitude{N(q_x, q_y=0)}\right]$ versus $q_x$ for
the case in which we only average over a $q_y$ segment of width $\dqy = 2
\pi (2 d +1)/L$ with $d=10$.  Thus the $q$-averaging brings out the underlying 
structure of the quasi-particle interference factor.

Finally, we note that even without such block averaging, one is still
able to see an image of the quasi-particle interference response
in an intensity plot of  
\begin{equation}
\left|\bar N(q, \omega)\right| =
\biggl|\frac{1}{\pi} Im\, \Lambda\, (q,\omega)\biggr|\ 
\frac{|\delta\epsilon(q)|}{\sqrt{N}}
\end{equation}
over the $(q_x,q_y)$ plane. In Fig.~20 we show a plot of 
$\left|\bar N(q, \omega)\right|$
for the same parameters that were used in Fig.~14. 
Here $|\delta\epsilon(q)|$ is obtained
from Eq.~(A1) with one realization of an impurity
concentration $n_i=0.01$. Comparing this figure with 
Fig.~14, one can see that while the structure is noisy, the key features
remain clearly visible.

\section{Quasi-particle Interference Response}

The quasi-particle interference response function for the 2D free electron gas 
\begin{equation}
\frac{1}{\pi} 
\, Im\, \Lambda (q, \omega)= \frac{1}{\pi}\, Im\, \int\, d^2x\, e^{iq\cdot x}
G_o(x, \omega)\, G_o(-x, \omega)
\label{aone}
\end{equation} 
can be directly evaluated using the expression for $G_o(x, \omega)$ given by
Eq.~(15).  Since the Hankel function $H^{(1)}_0= J_0 + i\, N_0$, we have
\begin{eqnarray} 
\frac{1}{\pi}\, Im\, \Lambda\, (q, \omega)& = & -2\pi\, N^2(0)\, \int d^2x
\, e^{i q \cdot x} J_0\, \left(k (\omega) r\right)
\, N_0\, \left(k (\omega) r\right)\nonumber\\
& = & \left\{\begin{array}{lccr}
8\pi\, N^2(0) & \frac{1}{q} & \frac{1}{\sqrt{q^2-4k^2 (\omega)}}
& q> 2k (\omega)\\
0 &&& q<2k (\omega) \end{array}\right. 
\label{atwo}
\end{eqnarray}
with $r=|x|$ and
\begin{equation}
k (\omega) = k_F\, \left(1+\frac{\omega}{\mu}\right)^{\frac{1}{2}}
\label{athree}
\end{equation}
We also note that
Carrying out the angular integration one has
\begin{equation}
Re \Lambda(q,\omega)= \left\{\begin{array}{lr}
\frac{8\pi N(0)^2}{q^2\sqrt{q^2-4k^2}} \ln 
\left |\frac{\sqrt{q^2-4k^2}+1}{\sqrt{q^2-4k^2}-1} \right| & q> 2k(\omega)\\
\qquad \frac{-16\pi N(0)^2}{q^2\sqrt{4k^2-q^2}}  \tan^-1 
\frac{q}{\sqrt{4k^2-q^2}} & q< 2k(\omega) \end{array} \right.
\end{equation}
Plots of the real and imaginary parts of $\Lambda(q\omega)$
are shown in Fig.~\ref{fnew}. As noted, in Born approximation
one only sees $Im \lambda(q,\omega)$.
However, when the scattering is stronger, giving
rise to a phase shift, both  $Re \lambda(q,\omega)$ and $Im \lambda(q,\omega)$
will be present in the quasi-particle interference response.
The imaginary part of $\Lambda$
can also be evaluated by noting that for $\omega > 0$,
\begin{eqnarray}
\frac{1}{\pi}\ Im\, \Lambda\, (q, \omega ) &=&
\frac{1}{\pi}\ Im\, \int \frac{d^2 k}{(2\pi)^2}
\ \frac{1}{\omega-\epsilon_{k+q} + i\delta}\ \frac{1}{\omega-\epsilon_k 
+ i\delta}
\nonumber\\
&=& 2\int \frac{d^2k}{(2\pi)^2}\ \delta\, (\omega - \epsilon_k) 
\ \frac{1}{\epsilon_{k+q} -\epsilon_k}
\label{afour}
\end{eqnarray}
Carrying out the angular integration one has
\begin{equation}
\int\limits^{2\pi}_0 \frac{d\phi}{2\pi}\ \frac{1}{\epsilon_{k+q}-\epsilon_k}
= \left\{\begin{array}{lr}
\frac{2m}{\sqrt{q^4-4k^2q^2}} & q^2> 4k^2\\
\qquad 0 & q^2< 4k^2 \end{array} \right.
\label{afive}
\end{equation}
so that
\begin{equation}
\frac{1}{\pi}\, Im\, \Lambda\, (q, \omega)  =  2N(0) 
\int\limits^{q^2/8m-\mu}_{-\mu}
d\epsilon_k\ \delta(\omega-\epsilon_k)\ \frac{2m}{\sqrt{q^4-4k^2q^2}} 
\label{asix}
\end{equation}
which gives Eq.~(\ref{atwo}) since $N(0)=m/2\pi$.

For the electron-phonon case in which the self-energy $\Sigma (\omega) = \Sigma_1
(\omega) + i\Sigma_2 (\omega)$ only depends on $\omega$, the
angular integral can be evaluated in the same way leading to 
\begin{equation}
\frac{1}{\pi}\, Im\, \Lambda\, (q, \omega) = 
2N(0) \int\limits^{q^2/8m-\mu}_{-\mu}
d\epsilon_k\ \frac{\frac{1}{\pi}
\ \Sigma_2 (\omega)}{\left(\omega-\epsilon_k - 
\Sigma_1 (\omega)\right)^2 + \Sigma^2_2 (\omega)}
\ \frac{2m}{\sqrt{q^4 - 4k^2 q^2}}.
\label{aseven}
\end{equation}
Here, as usual for the electron-phonon problem we have neglected vertex corrections.

For the case of a superconductor with scattering from a site charge potential
\begin{equation}
\Lambda\, (q, \omega)  =  \int\, \frac{d^2k}{(2\pi)^2}
\ \frac{(\omega + \epsilon_{k+q})\ (\omega + \epsilon_k) - \Delta_{k+q} 
\Delta_k}{\left(\omega^2 
- E^2_k + i\, \delta\right)\ \left(\omega^2-E^2_{k+q} + i\, \delta\right)}
\label{asevena}
\end{equation}
with $E_k = \sqrt{\epsilon^2_k + \Delta^2_k}$.  For a constant $s$-wave 
gap $\Delta_k=
\Delta_0$ and
\begin{equation}
\frac{1}{\pi}\ Im\, \Lambda\, (q, \omega) =  -\frac{2}{\pi}\, Im
\, \int\ \frac{d^2k}{(2\pi)^2}
\ \frac{\omega + \epsilon_k}{\omega^2-E^2_k
+ i\, \delta}\ \frac{1}{\epsilon_{k+q} - \epsilon_k}
\label{aeight}
\end{equation} 
Making use of Eq.~(\ref{afive}), we have
\begin{equation}
\frac{1}{\pi}\ Im\, \Lambda\, (q, \omega) = 2N(0) 
\int\limits^{q^2/8m-\mu}_{-\mu}
d\epsilon_k\, \delta\, \left(\omega^2-E^2_k\right)
\ \frac{2m}{\sqrt{q^4-4k^2q^2}}\ (\omega + \epsilon_k).
\label{anine}
\end{equation}
Carrying out the $\epsilon_k$ integration leads to the results given by 
Eqs.~(\ref{thirtyeight}) - (\ref{thirtynine}) in the text.

For a $d$-wave superconductor, we have
\begin{equation}
\frac{1}{\pi}\ Im\, \Lambda\, (q, \omega) = 2 \int \frac{dk^2}{(2\pi)^2}
\ \delta\, (\omega-E_k)\ \frac{1}{2E_k}\ \frac{(\omega + \epsilon_k)
\ (\omega+ \epsilon_{k+q}) - \Delta_k \Delta_{k+q}}{E^2_{k+q} - E^2_k}.
\label{aten}
\end{equation}
This integral can be approximately evaluated when 
$\omega < \Delta_0$ for the case
in which $\epsilon_k = k^2/2m - k^2_F/2m$ and
$\Delta_k=\Delta_0 \cos (2\theta)$.  For example, for ${\bf q}$ along the 45$^\circ$
direction with $q=q_x=q_y$ and $q \gtwid  k_F \omega/\Delta_0$, one
finds that
\begin{equation}
\frac{1}{\pi}\ Im\, \Lambda\, (q,\omega) \simeq 
 N^2 (0)\left( \frac{1}{N(0)\Delta_0}\right)
\ \frac{1}{k_F\sqrt{q}}\ \frac{1}{\sqrt{q-\frac{k_F\omega}{\Delta_0}}}
\label{aeleven}
\end{equation}
This corresponds to the contribution which comes from a momentum transfer that
connects the ends of one $\omega=\sqrt{\epsilon_k^2 + \Delta^2_k}$ 
contour (i.e. a ${\bf q}_7 (\omega)$ wave vector) similar 
to the case of the ellipse discussed in Section III. The enhancement factor
$(N(0)\Delta_0)^{-1}$ arises from the large curvature and resulting large
density of states at the contours ends.

Keeping $q$ along the 45$^\circ$ diagonal where $q=q_x=q_y$, there 
are additional square root peaks in $Im\, \Lambda (q, \omega)/\pi$ 
which arise when ${\bf q}$ connects two different contours.  
There is a ${\bf q}_3(\omega)$-like peak near which
\begin{equation}
\frac{1}{\pi}\, Im\, \Lambda (q,\omega) \simeq \frac{4\pi\, N^2(0)}{q}
\ \left(\frac{\omega}{2\Delta_0}\right)\ \frac{1}{\sqrt{k_F-q}}
\ \frac{1}{\sqrt{{\bf q}_3(\omega)-q}}
\label{athirteen}
\end{equation}
when
\begin{equation}
q<  q_3(\omega) = 2k_F \left(1-\frac{1}{2}
\ \left(\frac{\omega}{2\Delta_0}\right)^2\right).
\end{equation}
There is also a peak associated with
\begin{equation}
 q^\prime (\omega) = 2k_F \left(1+\frac{\omega}{\mu}\right)^{\frac{1}{2}}
\label{afourteen}
\end{equation}
where
\begin{equation}
\frac{1}{\pi}\, Im\, \Lambda (q, \omega) \simeq \frac{4\pi\, N^2(0)}{q}
\ \left(\frac{\omega}{2\Delta_0}\right)\ 
\frac{1}{\sqrt{q-q^\prime(\omega)}} \ \frac{1}{\sqrt{q - q_3(\omega)}}
\label{asixteen}
\end{equation}
when $q > q^\prime (\omega)$.  In the limit $\Delta_0\to0$, this last
expression becomes
\begin{equation}
\frac{1}{\pi}\, Im\, \Lambda (q, \omega) \simeq \frac{4\pi\, N^2(0)}{q
\sqrt{k_F}\ \sqrt{q-2k(\omega)}}
\end{equation}
which is the free electron result Eq.~(\ref{atwo}) for $\omega/\mu \ll 1$.

\acknowledgments

We would like to acknowledge useful discussions of the experimental FT-STM
measurements with J.C.~Davis, 
J.E.~Hoffman, A.~Kapitulnik, and K.~McElroy. We also
thank M.E.~Flatt\'e for discussions which rekindled our interest in this problem
and S.A.~Kivelson, P. Hirschfeld, R.L. Sugar and
L. Zhu, for many helpful and insightful discussions during the course of
our work.  This work was supported by the Department of Energy under Grant
\#DOE85-45197. We would also like to acknowledge support provided by the 
Yzurdiaga gift to UCSB.

\begin{figure}
\vspace{10mm}
\centerline{\epsfysize=7cm\epsfbox[130 225 480 570]{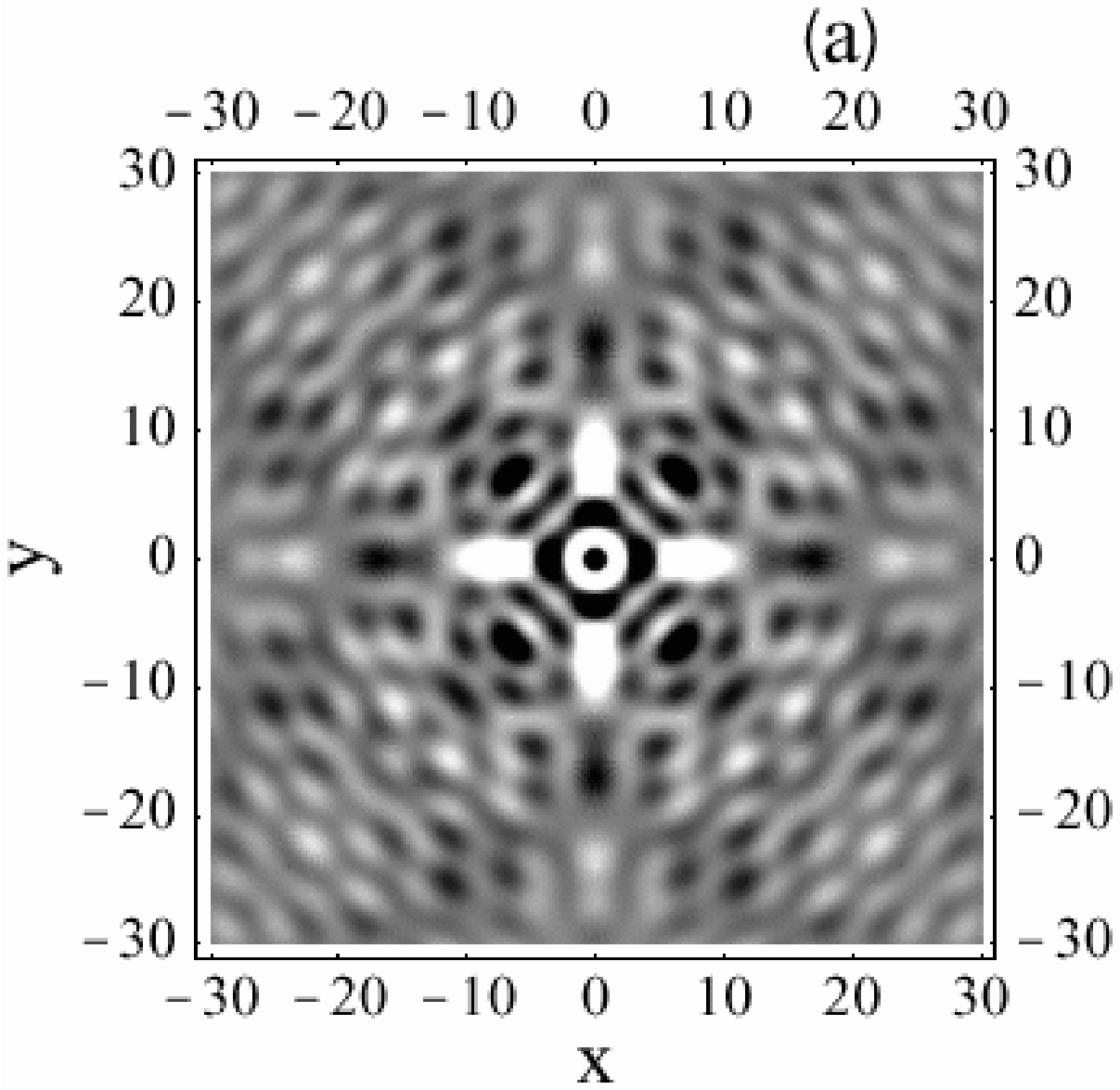}\hskip 0.2cm
\epsfysize=6cm\epsfbox[15 140 530 640]{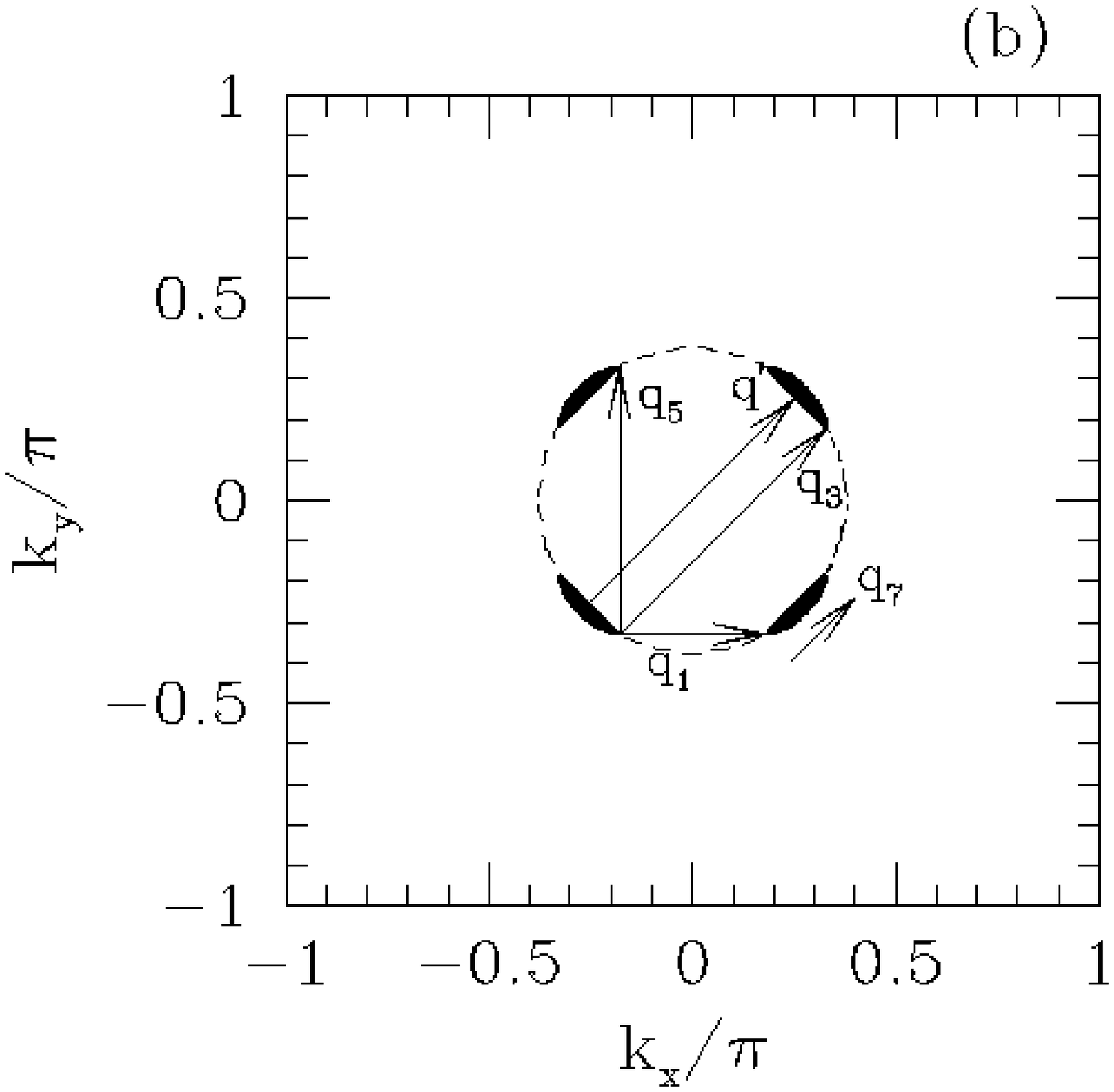}}
\centerline{\epsfysize=7cm\epsfbox[130 225 480 570]{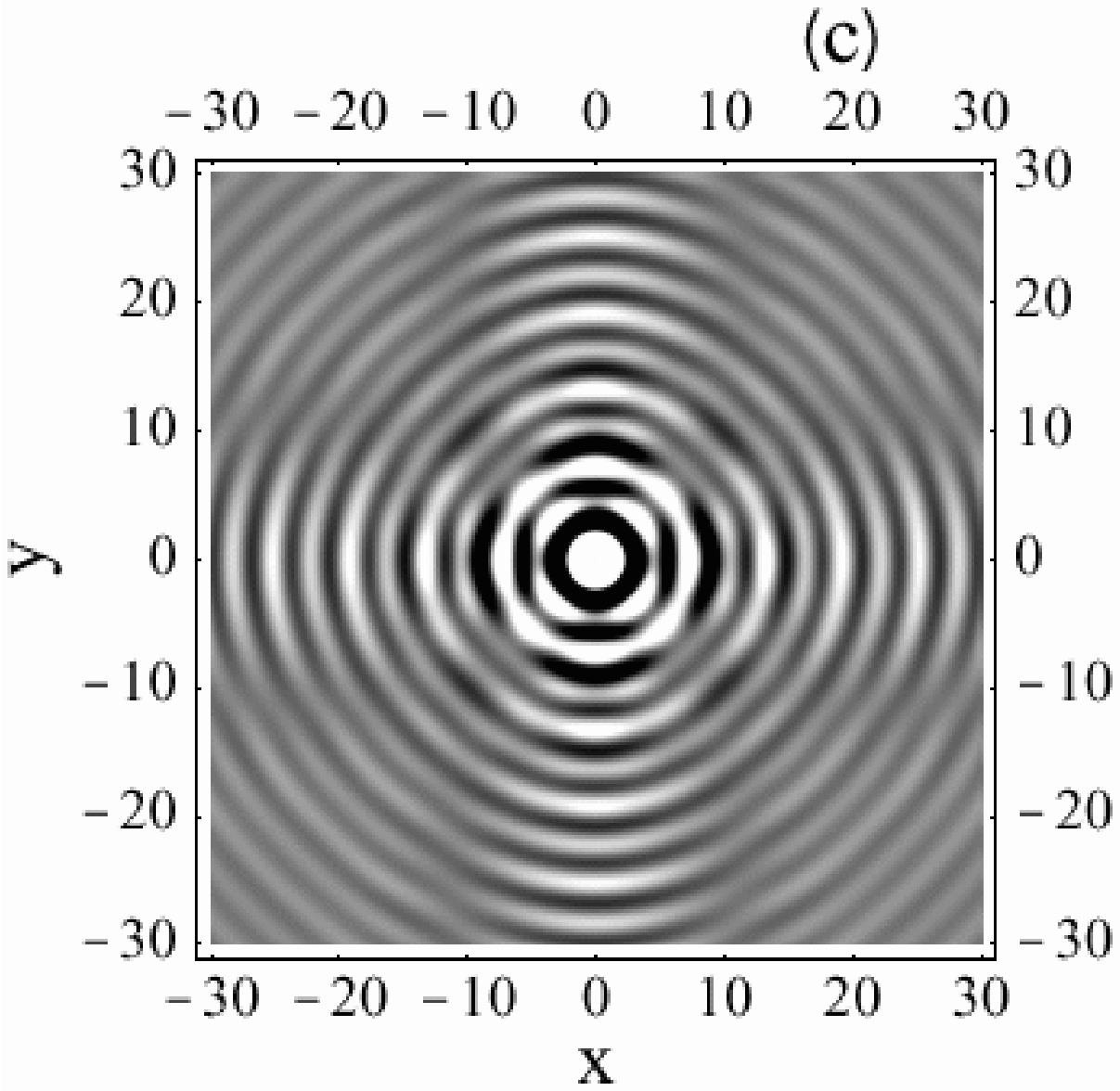}\hskip 0.2cm
\epsfysize=6cm\epsfbox[15 140 530 640]{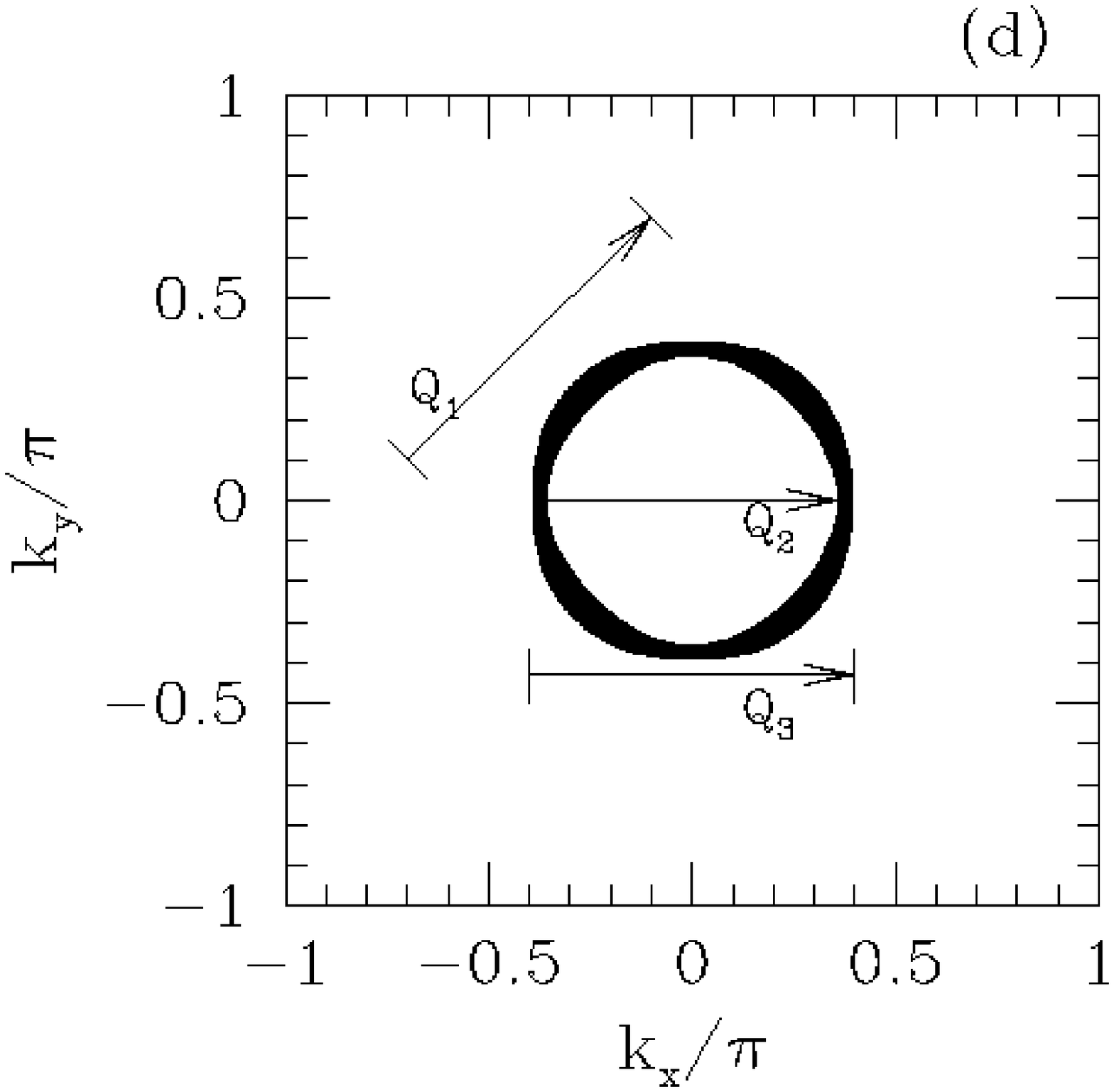}}
\caption{(a) The local tunneling density of states $N(x,\omega)$ 
due to scattering from a weak potential at the origin \cite{BFS93},
for (a) $\omega=0.5\Delta_0$ and (c)  $\omega=1.1\Delta_0$. 
The length scale is set by $k_F^{-1}$.
These results were obtained for a cylindrical Fermi surface with a
$d$-wave gap $\Delta(\theta)= \Delta_0\cos (2\theta)$.
The contours of the solid regions in (b) and (d) 
correspond to the points where $\omega=\sqrt{\epsilon^2_k + \Delta^2(\theta)}$ 
for $\omega=0.5\Delta_0$ and $1.1\Delta_0$, respectively. 
The dashed line in (b) is the Fermi surface
for the non-interacting system.
The ${\bf q}_\alpha(\omega)$ wave vectors in (b), 
introduced in Ref.~\cite{McE02}, connect the tips
of various contours and ${\bf q}^\prime(\omega)$ connects what would be 
a diagonal nesting vector for the normal state Fermi surface.
The ${\bf Q}_\alpha(\omega)$ wave vectors shown in (d) are the 
relevant nesting vectors when $\omega>\Delta_0$.} 
\end{figure}

\begin{figure}[h]
\includegraphics[width=0.80\textwidth]{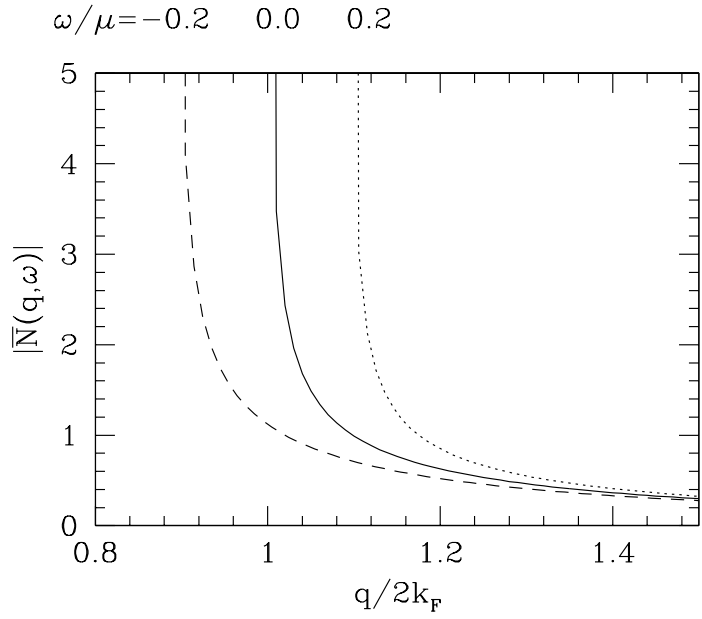}
\caption{Plot of the normalized $|N(q,\omega)|$ 
versus $q$ for a free electron gas for various values of $\omega/\mu$.}
\end{figure}

\begin{figure}[h]
\includegraphics[width=0.80\textwidth]{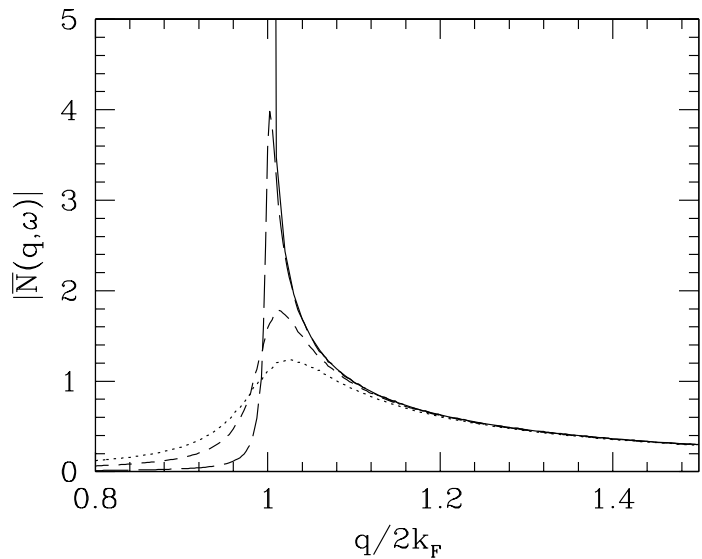}
\caption{Plot of the normalized mean-free path corrected $|N(q, \omega)|$ 
versus $q$ for an electron gas
with $\omega=0$. Here  $\Gamma = (\ell \, k_F)^{-1}$: 
$\Gamma=0$ (solid), $\Gamma=0.01$ (long dashed), $\Gamma=0.05$ (dashed),
$\Gamma=0.1$ (dotted).}
\end{figure}
\begin{figure}[h]
\includegraphics[width=0.80\textwidth]{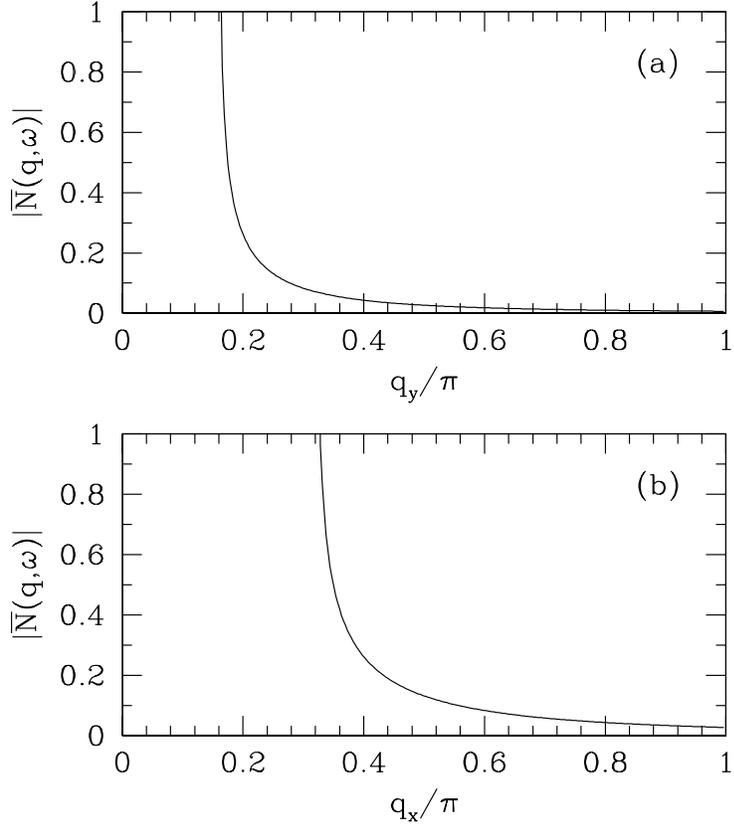}
\caption{Plot of the normalized $|N(q, \omega)|$ versus $q$,
Eq.~(\ref{twentyseven}), for an elliptical Fermi surface with $\gamma^2=
m_x/m_y=9$ and $\omega=0$.
In (a) $q_x=0$ and in (b) $q_y=0$}
\end{figure}
\begin{figure}[h]
\includegraphics[width=0.80\textwidth]{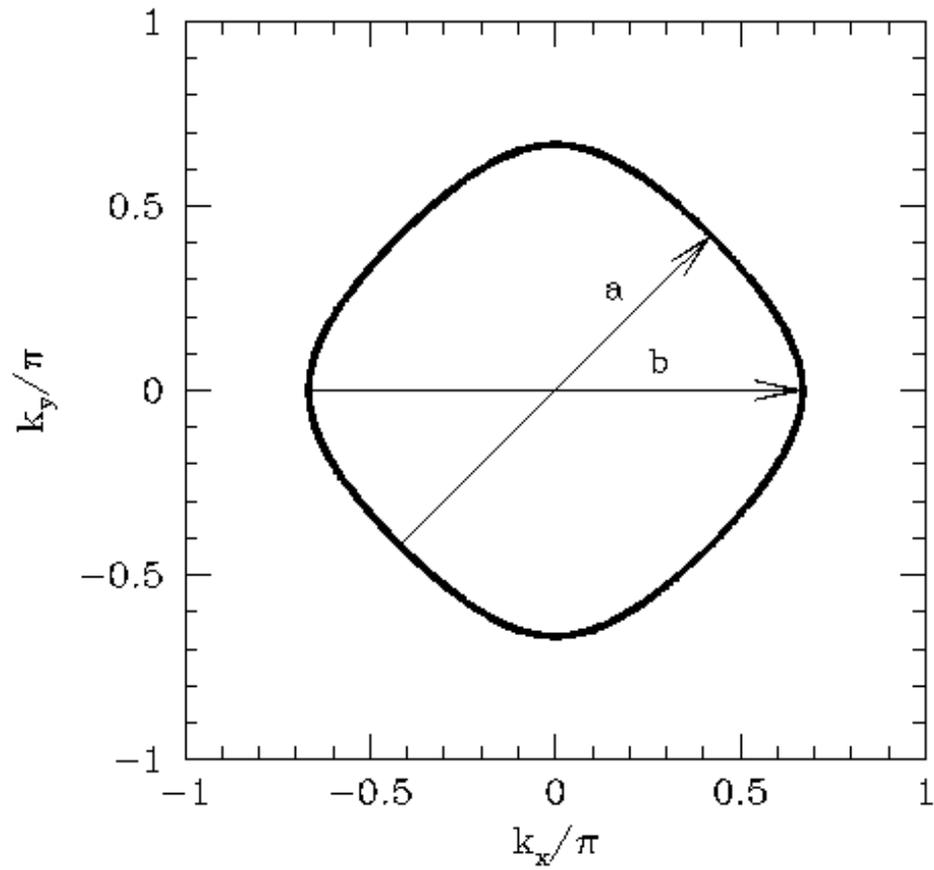}
\caption{Fermi surface of a near-neighbor tight-binding band, 
Eq.~(\ref{tb}), with $\mu/t=-1.0$ corresponding to a site 
filling $\langle n\rangle \approx 0.31$.}
\end{figure}
\begin{figure}[h]
\includegraphics[width=0.80\textwidth]{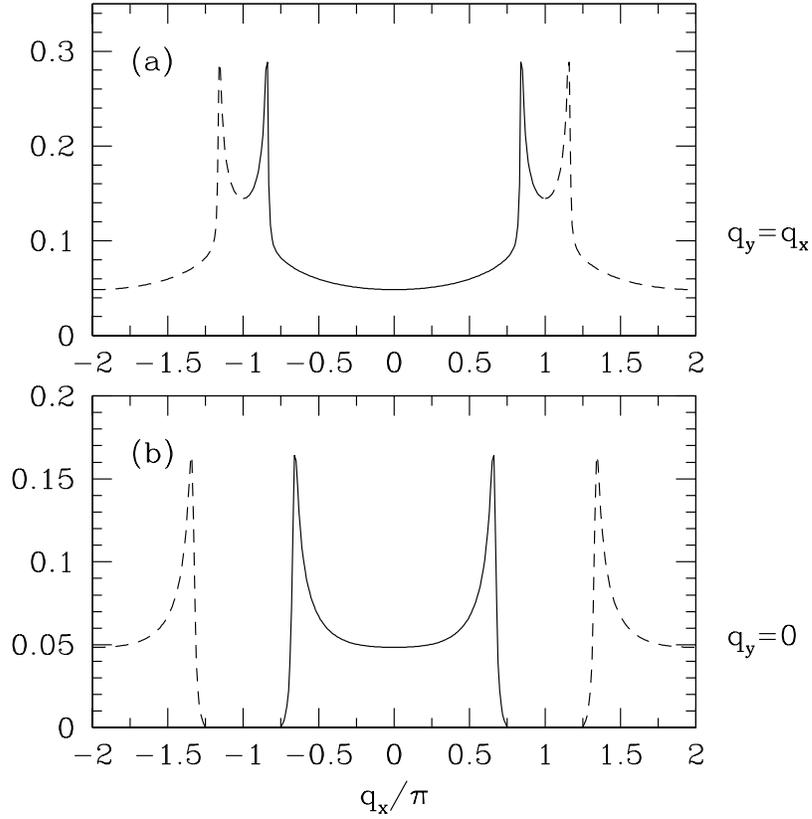}
\caption{$|Im\, \Lambda (q, \omega)/\pi|$ for the tight-binding
band, Eq.~\ref{thirty}, with $t=1$ 
versus $q$ for $\omega=0$ and (a) the diagonal cut-a and (b) 
the horizontal cut-b indicated in Fig.~5. The dashed curves
show the results in the second Brillouin zone.  
For the cut-a, $2k_F^{xy} \simeq .8\pi$
so the solid curve in the first Brillouin looks similar to the free electron
response.  However, for cut-b, $2k_F^{x} \simeq 1.3\pi$ 
so that the dashed curve in (b)
looks similar to the free electron response and the solid curve in the first
Brilloin zone results from folding the dashed curve back into the first zone.}
\end{figure}
\begin{figure}[h]
\includegraphics[width=0.80\textwidth]{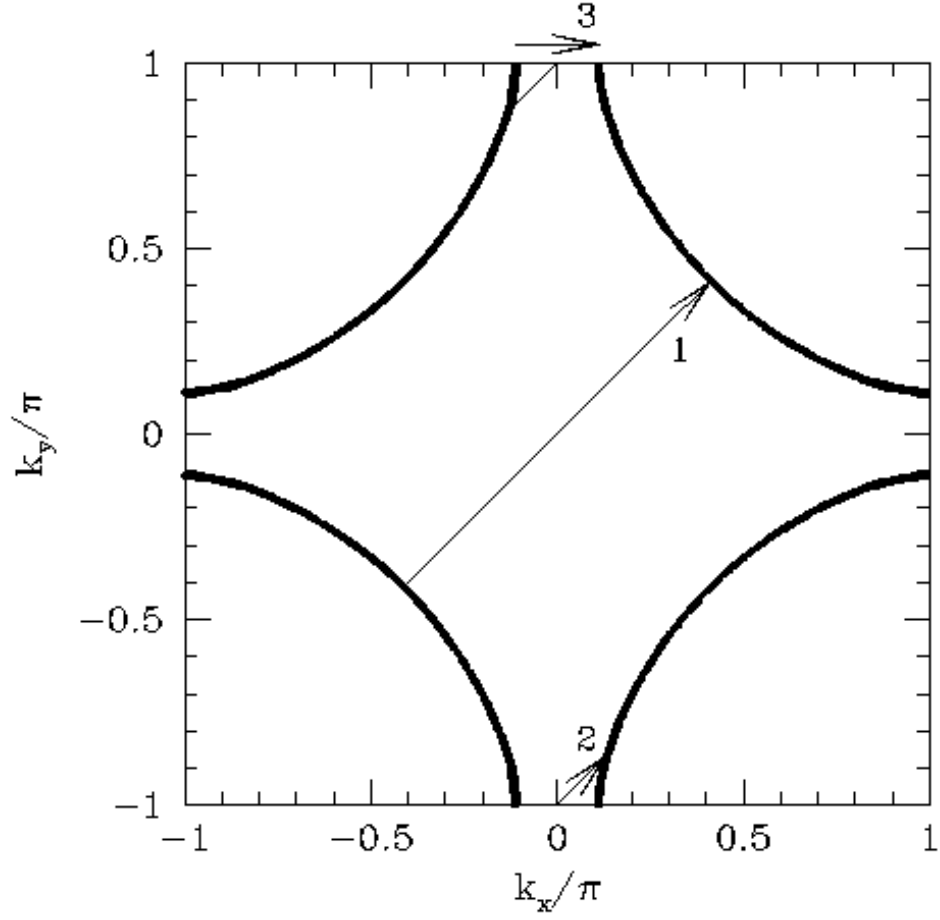}
\caption{Fermi surface of a tight-binding band with near- and
next-near-hopping,
Eq.~(\ref{tb}), with $t^\prime/t=-0.3$ and $\mu/t=-1.0$.}
\end{figure}
\begin{figure}[h]
\includegraphics[width=0.80\textwidth]{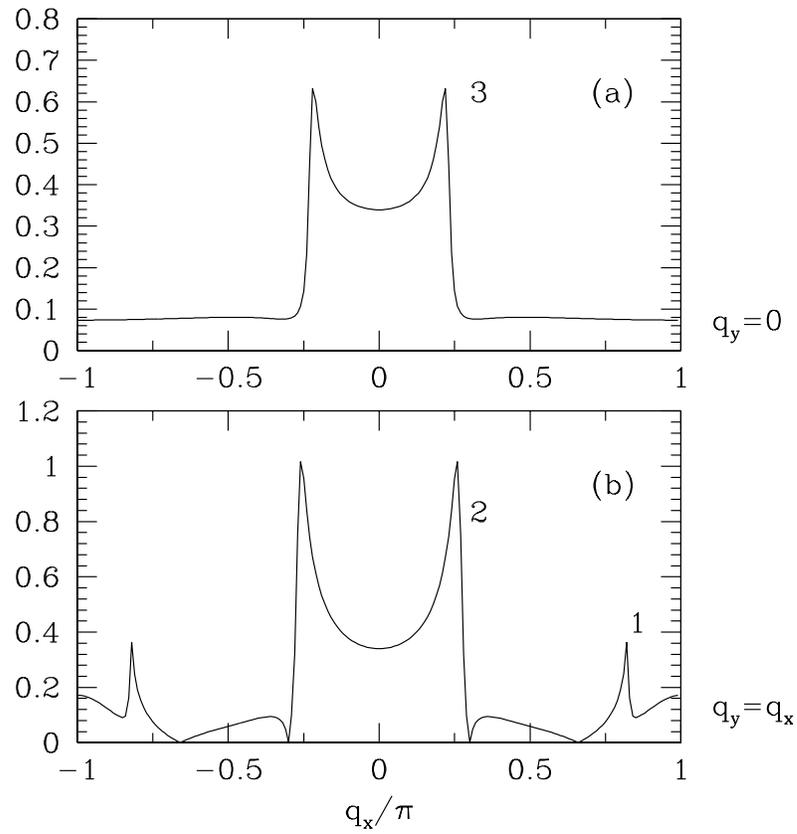}
\caption{$|Im\, \Lambda (q, \omega)/\pi|$ versus $q$ for $\omega=0$ for the
tight-binding Fermi surface shown in Fig.~7. (a) $q_y=0$ and (b) $q_x=q_y$.}
\end{figure}
\begin{figure}[h]
\includegraphics[width=0.45\textwidth]{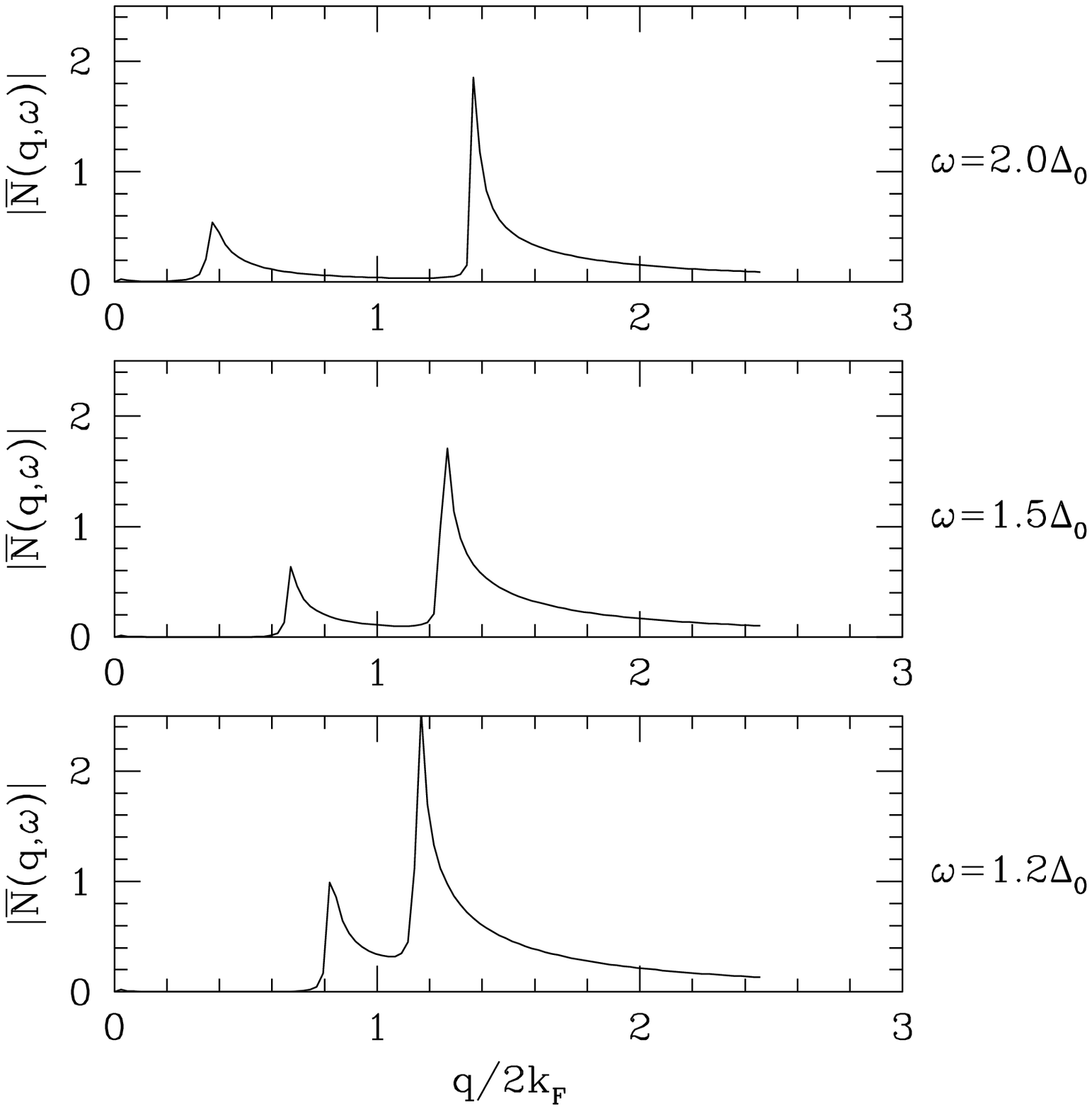}
\includegraphics[width=0.45\textwidth]{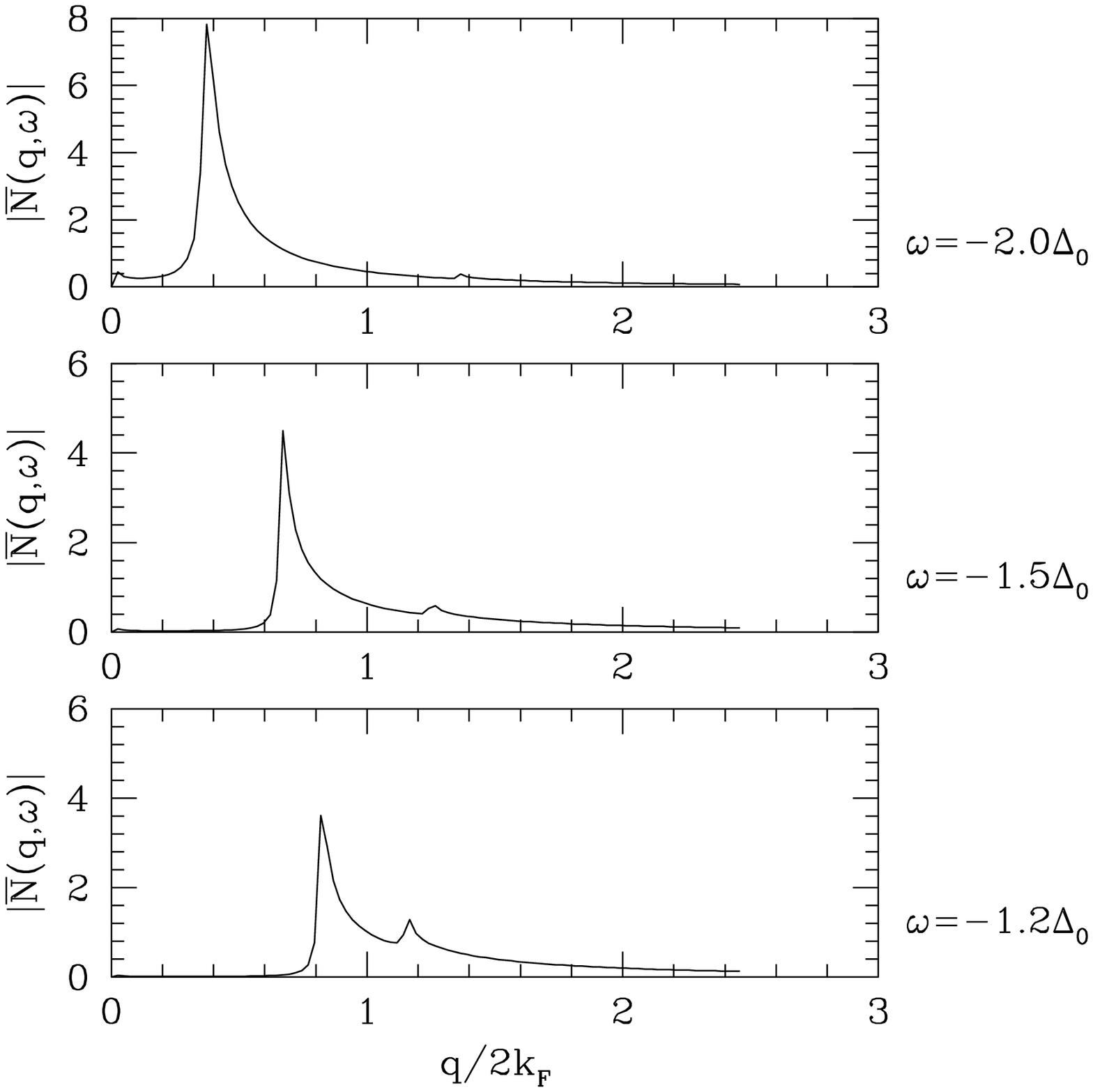}
\caption{Plot of the normalized $|N(q, \omega)|$ 
versus $q$ for an $s$-wave superconductor with a
cylindrical Fermi surface and $\Delta/\mu=0.5$. (a) Results for positive
values of $\omega=eV$ and (b) for negative values of $\omega$.} 
\end{figure}
\begin{figure}[h]
\includegraphics[width=0.45\textwidth]{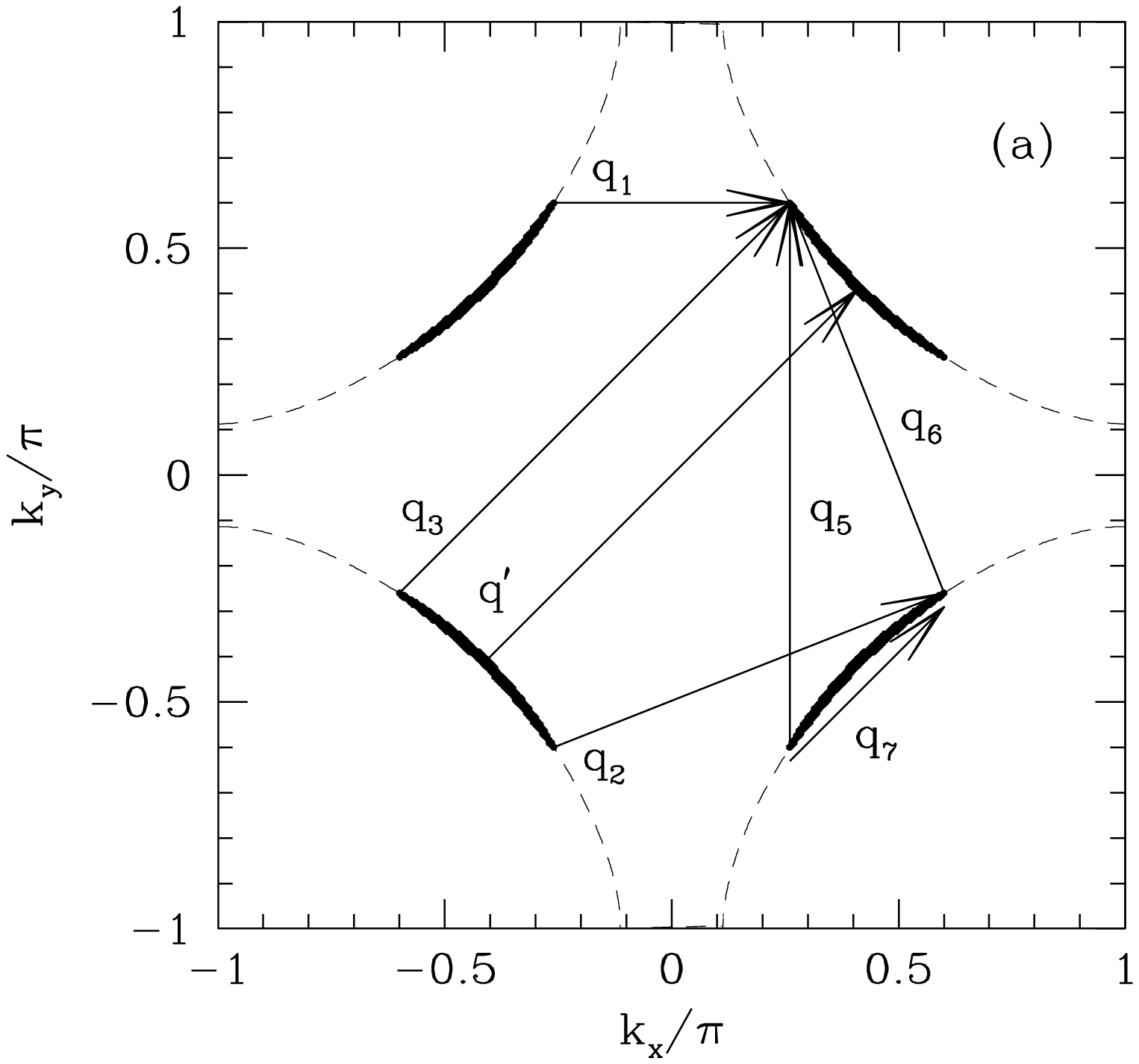}
\includegraphics[width=0.45\textwidth]{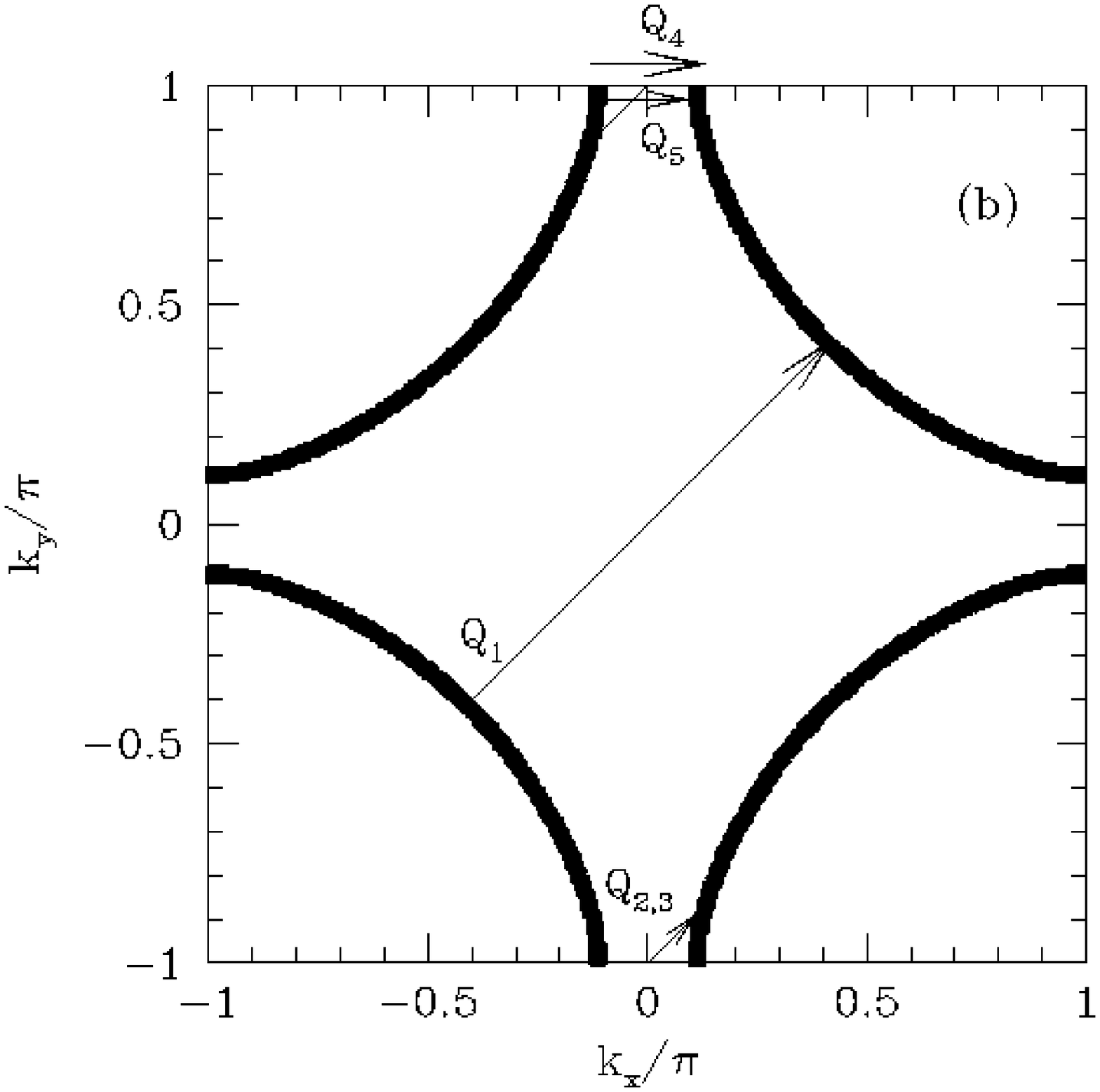}
\caption{Characteristic wave vectors for a d-wave superconductor
with a tight-binding band, Eq.~(\ref{tb}), with
$t^\prime/t=-0.3$, $\mu/t=-1.0$, and $\Delta_0=0.1t$. The
contours of the solid regions correspond to the points
where $\omega=\sqrt{\epsilon_k^2+\Delta_k^2}$ with (a)
$\omega=0.5\Delta_0$, and (b) $\omega=1.1\Delta_0$.
The dashed line in (a) shows the Fermi surface of the
non-interacting system. In (b) $Q_3$ and $Q_5$ ($Q_4$ and $Q_2$)
connect the innermost (outermost) surfaces while
$Q_1$ connects the outermost surfaces.}\label{dwave}
\end{figure}

\begin{figure}[h]
\includegraphics[width=0.45\textwidth]{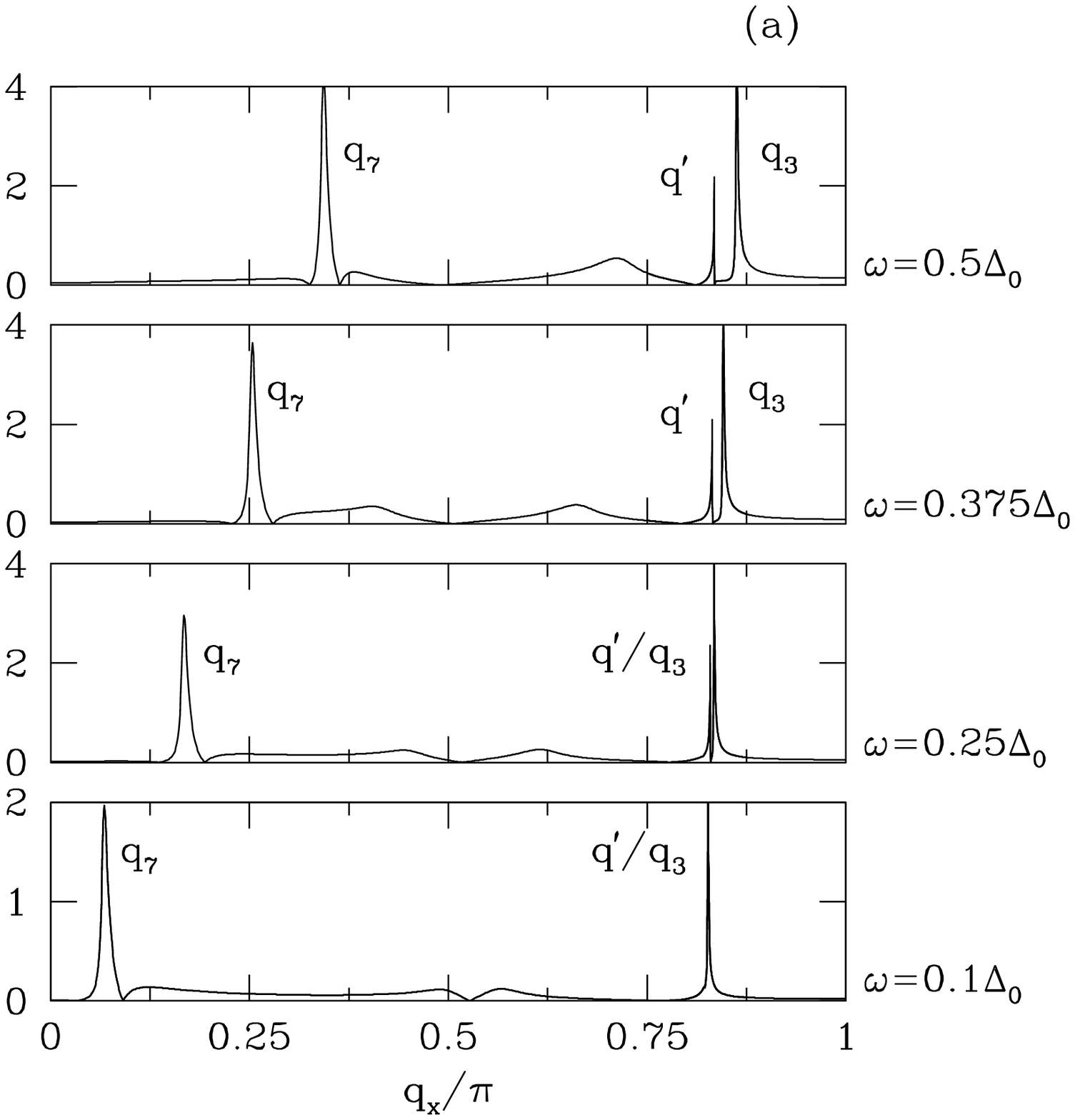}
\includegraphics[width=0.45\textwidth]{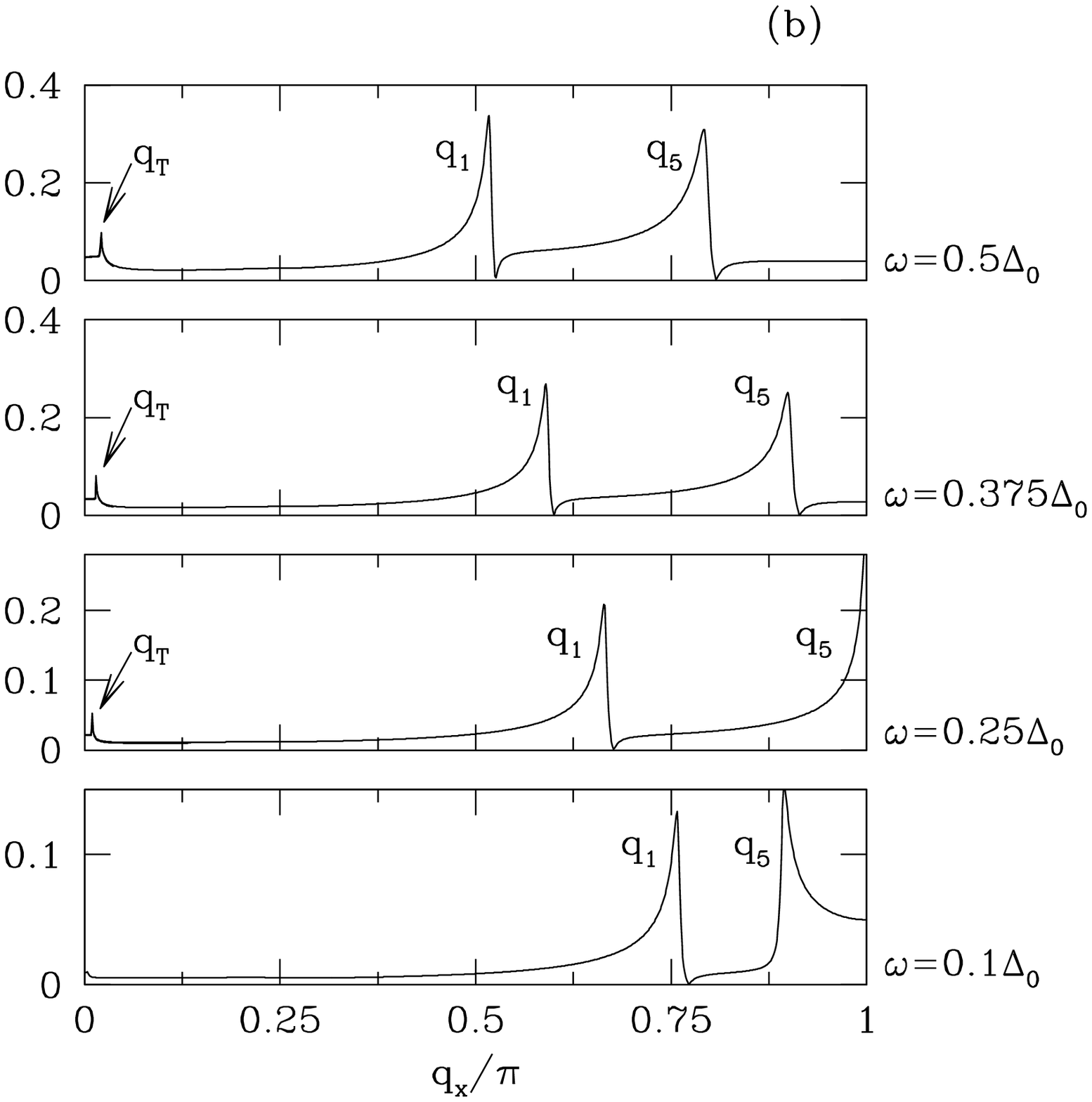}
\caption{$|Im\, \Lambda (q, \omega)/\pi|$ versus $q$ for a $d$-wave 
superconductor with a tight-binding band given by Eq.~(\ref{tb}) with
$t^\prime/t=-0.3$, $\mu/t=-1.0$, $\Delta_0=0.1t$.  
Here (a) shows results
for a diagonal cut for $q_x=q_y$ and (b) shows results for a
horizontal $q_y=0$ cut. The $q_\alpha$ and $q^\prime$ vectors are shown in 
Fig~\protect\ref{dwave}(a).}\label{dwave1}
\end{figure}
\begin{figure}[h]
\includegraphics[width=0.8\textwidth]{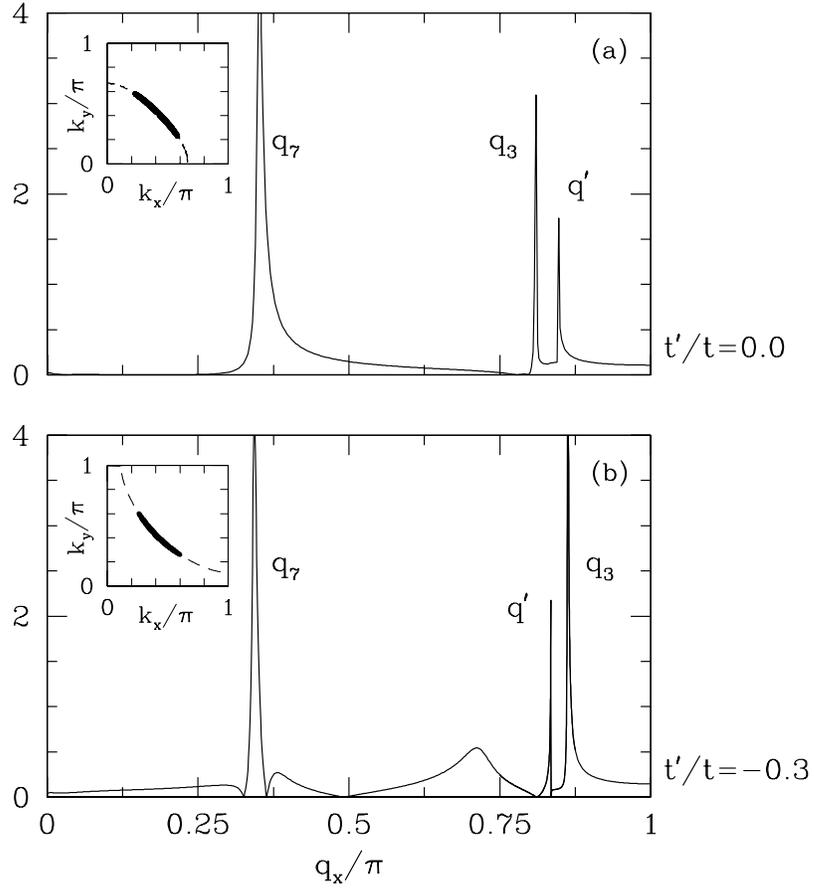}
\caption{$|Im\,\Lambda (q,\omega)/\pi|$ versus $q$ for a diagonal cut in
which $q_x=q_y$. (a) shows results for a tight-binding band with $t^\prime/t=
0$ and (b) shows results for $t^\prime/t = -0.3$. 
$\omega=0.5\Delta_0$ 
and the other parameters are the same of Fig.~\ref{dwave1}.}
\end{figure}
\begin{figure}
\includegraphics[width=0.8\textwidth]{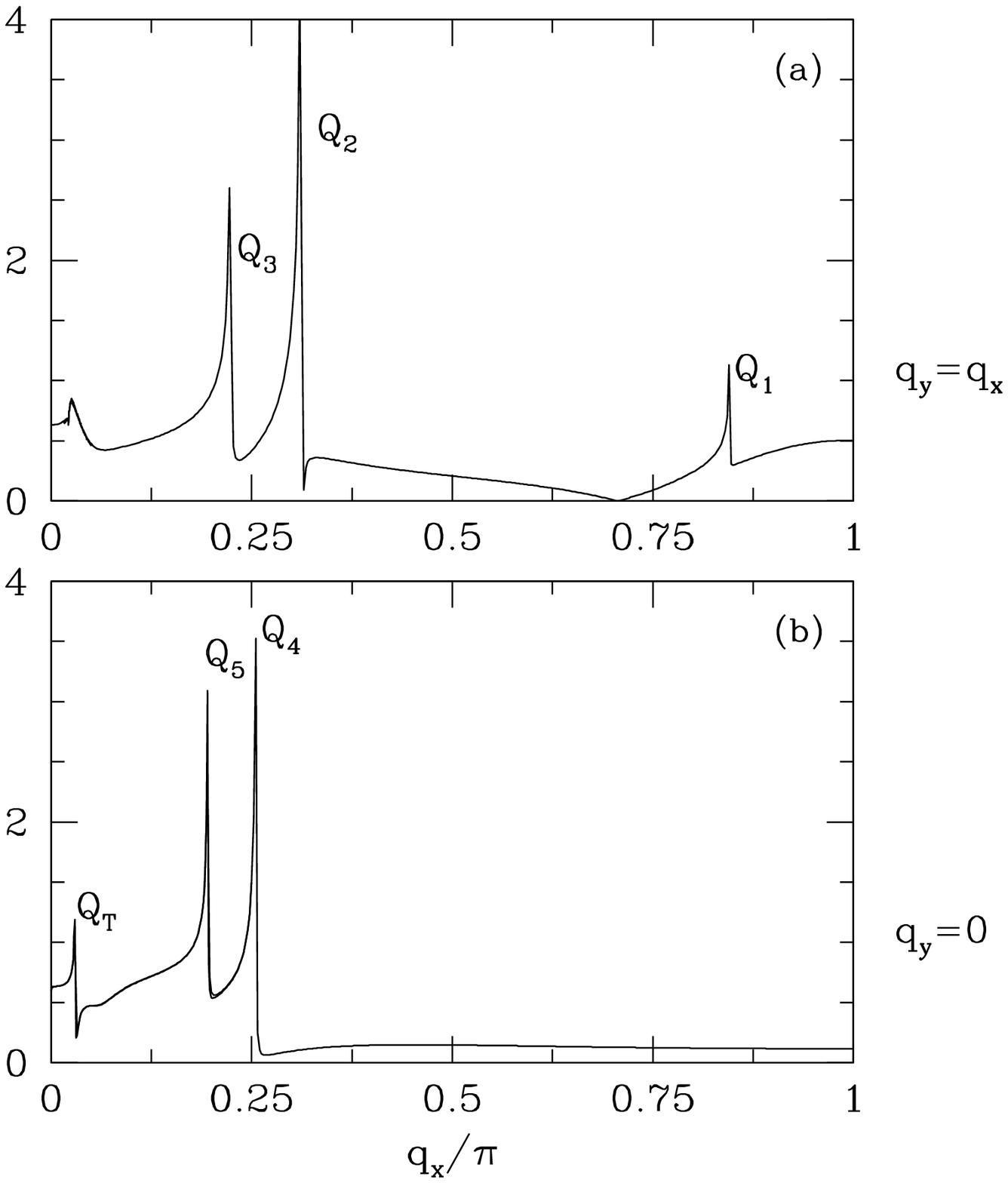}
\caption{$|Im\, \Lambda (q, \omega)/\pi|$ versus $q$ for a $d$-wave
superconductor with a tight-binding band given by Eq.~(\ref{tb}) with
$t^\prime/t=-0.3$, $\mu/t=-1.0$, $\Delta_0=0.1t$, and $\omega=1.1\Delta_0$.
Here (a) shows results
for a diagonal 45$^\circ$ cut for $q_x=q_y$ and (b) shows results for a
horizontal $q_y=0$ cut. The $Q_\alpha$ vectors are shown in
Fig~\ref{dwave}(b).}
\end{figure}

\begin{figure}
\includegraphics[width=0.8\textwidth]{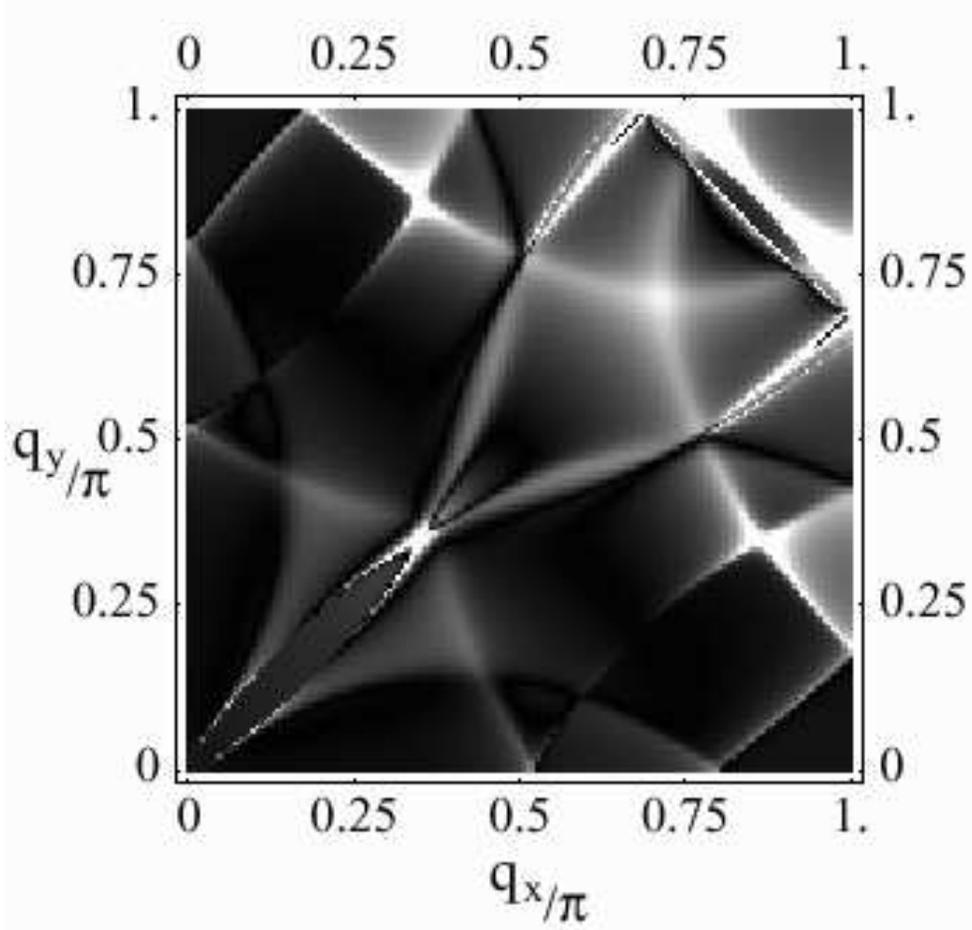}
\caption{Intensity map of the quasi-particle interference
response $|Im\, \Lambda (q, \omega)/\pi|$ for a $d$-wave
superconductor plotted over the first $(q_x,q_y)$
quadrant. Here, the band structure is given by Eq.~(\ref{tb}) with
$t^\prime/t=0.3$ and $\mu=-1.0$, $\Delta_k$ is given
by Eq.~(38) and the bias is $\omega=0.5\Delta_0$.}
\end{figure}

\begin{figure}
\includegraphics[width=0.8\textwidth]{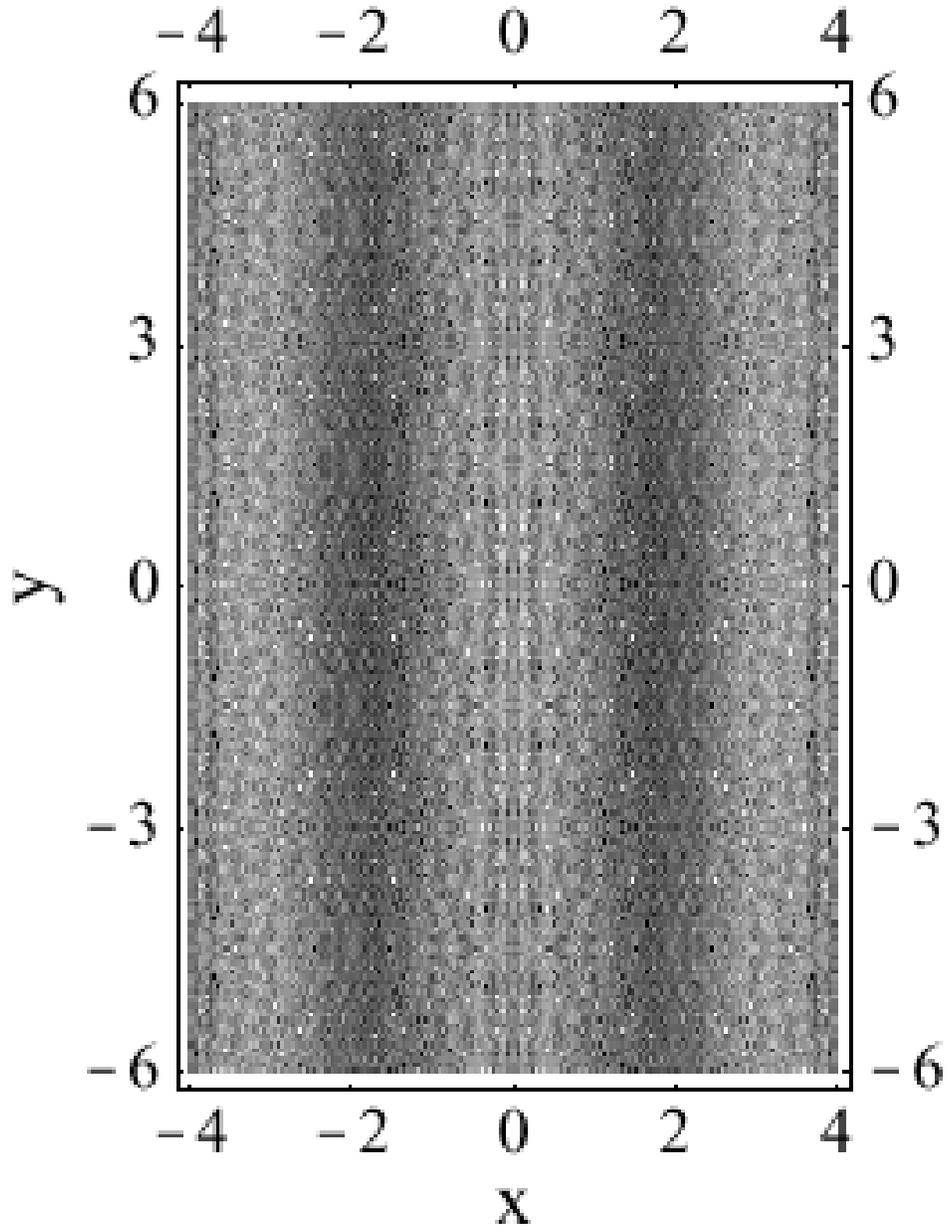}
\caption{Ripples in $N(x,\omega)$ when there is an ordered array
of scattering centers running along the $y$-axis. Here the spacing between 
the stripes correspond to 4 lattice sites and $\omega=0.5\Delta_0$.}
\end{figure}

\begin{figure}
\includegraphics[width=0.8\textwidth]{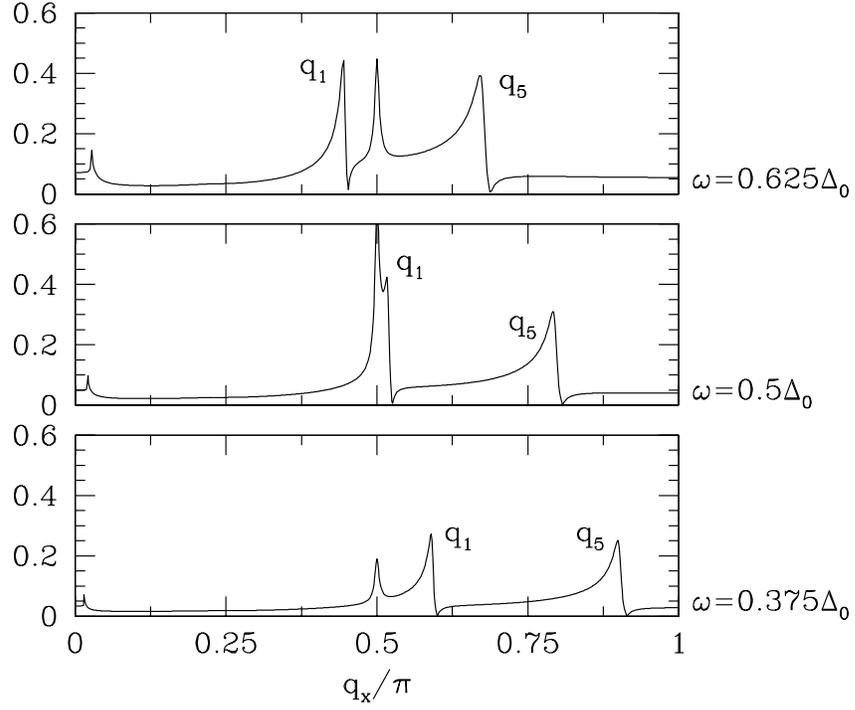}
\caption{Structure in $\left|\bar N(q, \omega)\right|$ versus 
$q_x$ for random impurities and an ordered array of stripes 
separated by 4 lattice spacings. The quasi-particle interference peaks
associated with $q_1(\omega)$ and $q_5(\omega)$, previously
shown in Fig.~11(b), are seen along with a peak at $q\simeq 0.5\pi$ which
arises from the striped array of scattering centers. }
\end{figure}

\begin{figure}[t]
\includegraphics[width=0.8\textwidth]{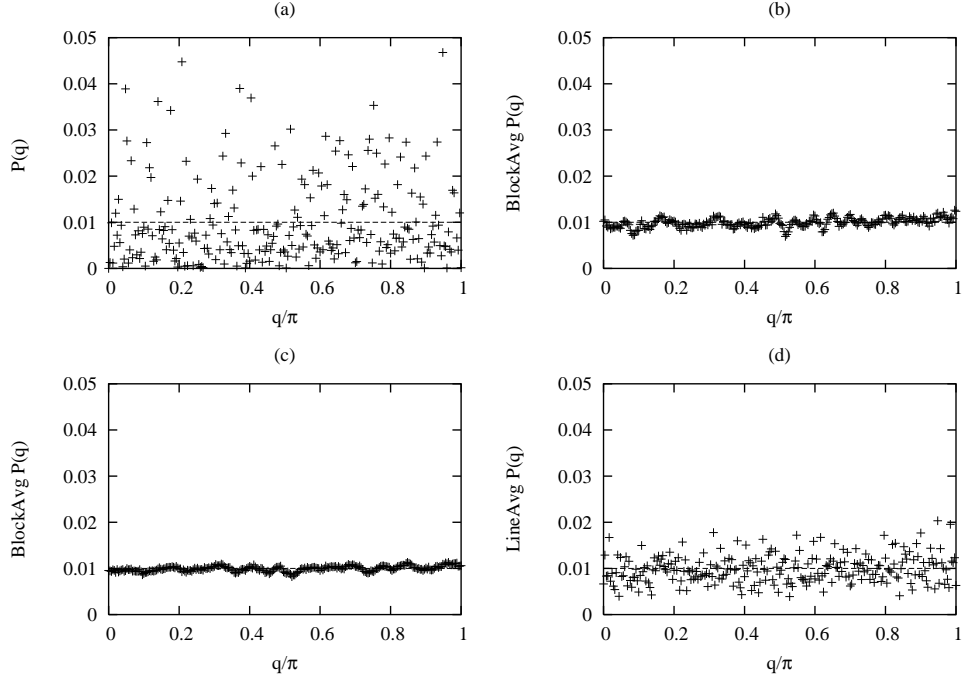}
\caption{Power spectrum, as defined in Eq.~(\ref{eqn:powspec}) for a random 
configuration of impurities with concentration $n_i=0.01$.  Here (a) shows 
the unaveraged results, (b) and (c) show the average over 
blocks of size $\Delta q_x = \Delta q_y = 2 \pi (2d+1)/L$ with $d=5$ and $d=10$ 
respectively, and (d) shows the average over lines of length
$\Delta q_y =2\pi (2d+1)$ with $d=10$.}
\end{figure}
\begin{figure}[t]
\includegraphics[width=0.8\textwidth]{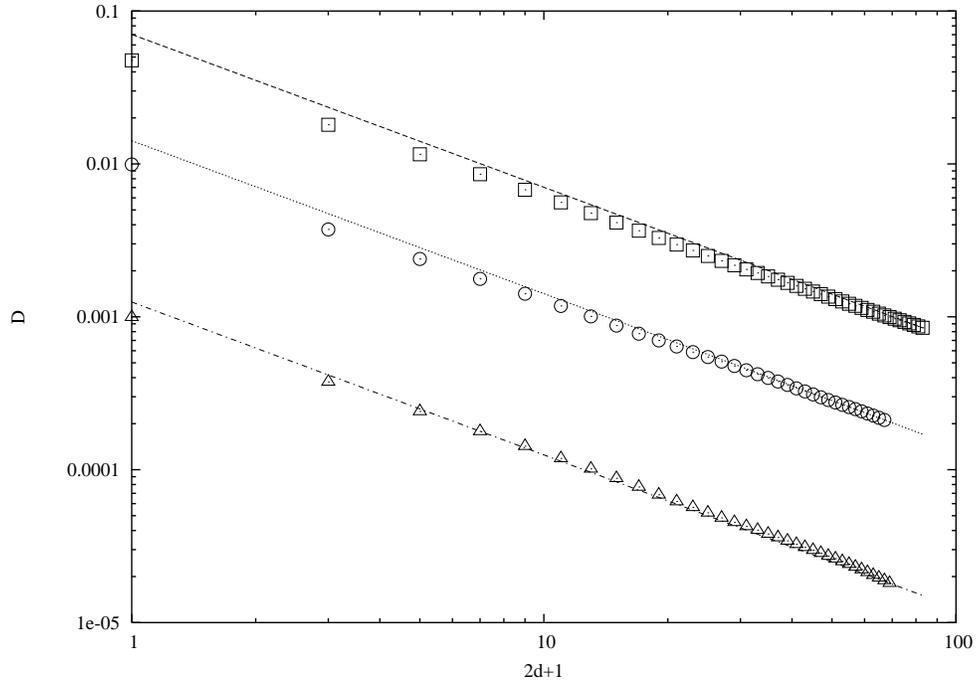}
\caption{RMS deviation of $\bar{P}(q)$ from $n_i$ as defined in 
Eq.~(\ref{eqn:deviation}).  The straight lines represent the asymptotic 
behavior in which $D \sim \alpha n_i/(2d+1)$.}
\end{figure}
\begin{figure}[t]
\includegraphics[width=0.8\textwidth]{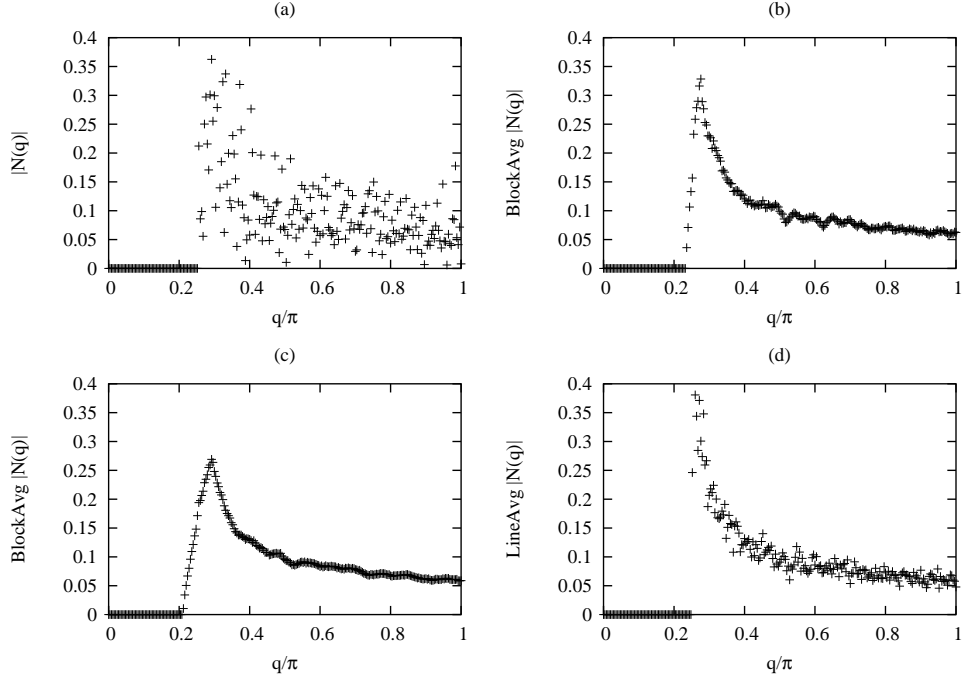}
\caption{The magnitude of the Fourier transform of the local tunneling 
density of states, as defined in Eq.~(\ref{eqn:nq}) for a random 
configuration of impurities with concentration $n_i=0.01$.  Here (a) shows 
the unaveraged results, (b) and (c) show the average over 
blocks of size $\Delta q_x = \Delta q_y = 2 \pi (2d+1)/L$ with $d=5$ and $d=10$ 
respectively, and (d) shows the average over lines of length
$\Delta q_y =2\pi (2d+1)$ with $d=10$.}
\end{figure}

\begin{figure}[t]
\includegraphics[width=0.8\textwidth]{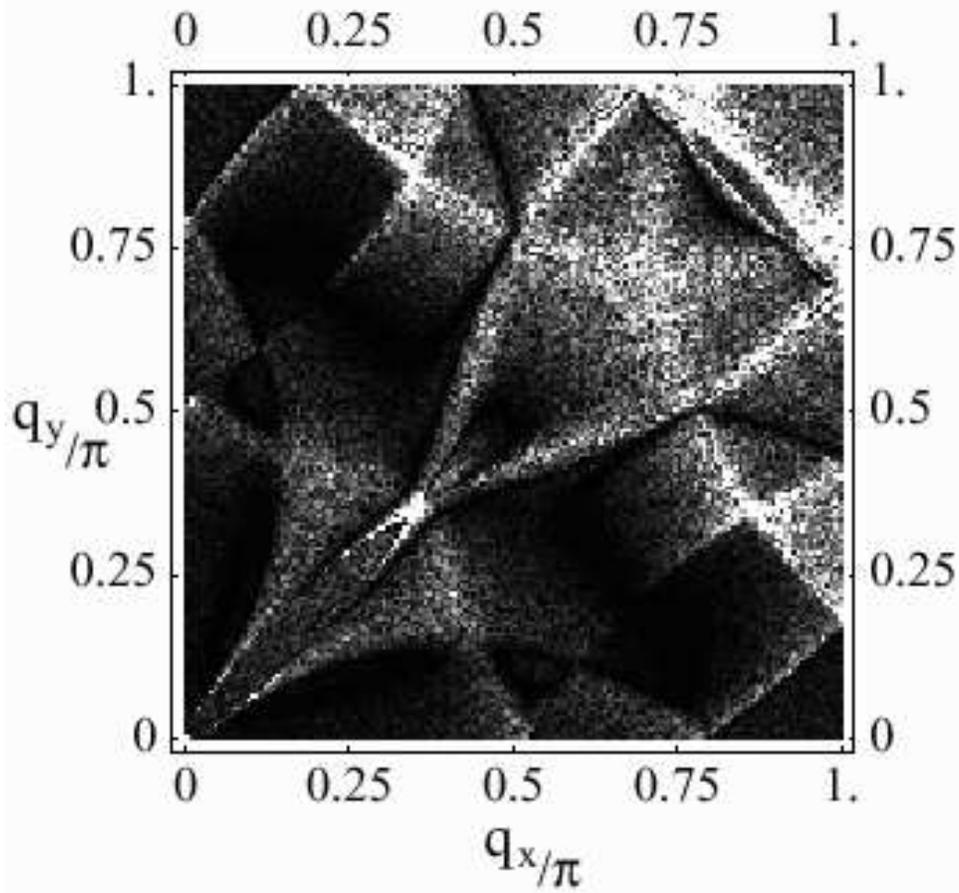}
\caption{Intensity map of 
$\left|\bar N(q, \omega)\right|$
for the same parameters that were used in Fig.~14. Here $|\delta\epsilon(q)|$ is obtained
from Eq.~(A1) with one realization of an impurity
concentration $n_i=0.01$.}
\end{figure} 

\begin{figure}[t]
\includegraphics{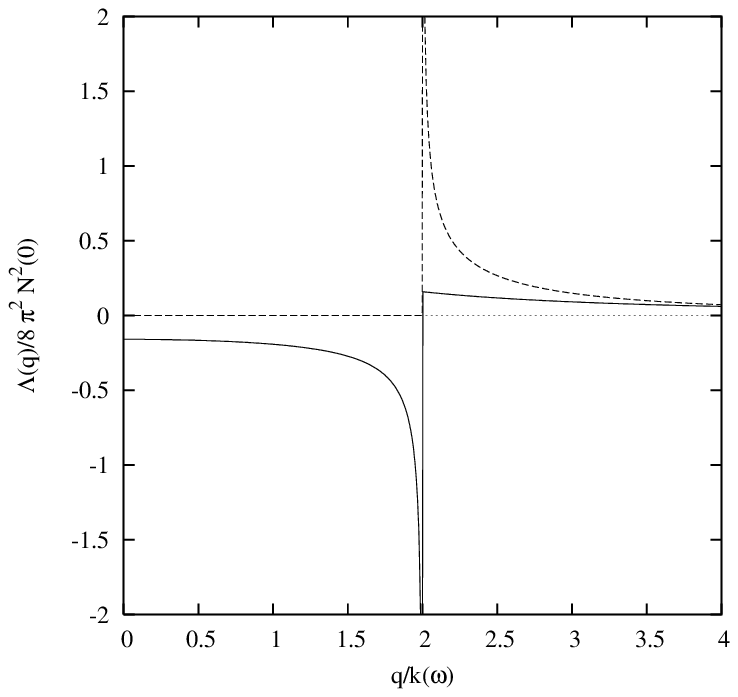}
\caption{\label{fnew}
The real (solid) and imaginary (dashed) part of the quasi-particle
interference response versus $q$ for a non-interacting 2D electron gas.}
\end{figure}

\printfigures


\begin{thebibliography}{99}


\bibitem{Hof02} J.E.~Hoffman, K.~McElroy, D.-H.~Lee, K.M.~Lang, H.~Eisaki,
S.~Uchida, and J.C.~Davis,  Science, {\bf 295}, 466 (2002).

\bibitem{How02} C.~Howald, H.~Eisaki, N.~Kaneko, and A.~Kapitulnik, e.print
cond-mat/0201546.

\bibitem{McE02} 
K.~McElroy, R.W.~Simmonds, J.E.~Hoffman, D.-H.~Lee, J.~Orenstein,
H.~Eisaki, S.~Uchida, and J.C.~Davis, to appear in Nature, March 2003.

\bibitem{BFS93} J.M.~Byers, M.E.~Flatt\'e, and D.J.~Scalapino, 
\prl {\bf 71}, 3363 (1993).

  
\bibitem{MB01} There have been various discussions of the effect of the tunneling
matrix elements on the relationship between the conductance $dI(V, x)/dV$ measured
on the $BiO_2$ layer and the local density of the states of the underlying $CuO_2$
layer. See for example I.~Martin and A.V. Balatsky, Physica C {\bf 357--360},
46 (2001).  Here, for most of our present discussion, we assume 
that it is the local density of states $N(x, \omega)$ at the $Cu$ 
site that is probed. 

\bibitem{FB99} M.E.~Flatt\'e and J.M.~Byers, Solid State Physics {\bf 52}, 
137 (1999).

\bibitem{Din96} H.~Ding, J.C.~Campuzano, M.~Randeira,
A.F.~Bellman, T.~Yokoya, T.~Takahashi, T.~Mochiku, and K. Kadowaki,
\prb {\bf 54}, R9678 (1996).

\bibitem{WL02} Q.-H.~Wang and D.-H.~Lee, 
e.print cond-mat/0205118.

\bibitem{Kivup} A calculation of $Im\, \Lambda (q,\omega)/\pi$ for the free
electron case as well as a discussion of screening corrections of the scattering
potential are given in
S.A.~Kivelson, E.~Fradkin, V.~Oganesyan, I.P.~Bindloss, J.M.~Tranquada,
A.~Kapitulnik, and C.~Howald, ``How to Detect Fluctuating Order in the High
Temperature Superconductors'', e. print cond-mat/0210683.

\bibitem{thnkKivelson} We thank S.~Kivelson for discussing this point with us.

\bibitem{ref2} Here we have considered only scattering by a site charge
potential, one
could also consider scattering from a variation in the gap or spin degrees of freedom
by changing the coherence factors in $\Lambda (q, \omega)$.

\bibitem{Tom65} W.J.~Tomasch, \prl {\bf 15}, 672 (1965). 

\bibitem{PDDH02} D.~Podolsky, E.~Demler, K.~Damle, and B.I.~Halperin, 
e.print cond-mat/0204011; A. Polkovnikov, S. Sachdev, and M. Vojta, cond-mat/0208334.




\end{thebibliography}
\end{document}